\documentclass[aps,pre,floatfix,amsmath]{revtex4-1}

\usepackage[english]{babel}
\usepackage{amsmath}
\usepackage{amssymb}
\usepackage{graphicx}

\begin{document}

\title{Record Statistics for Multiple Random Walks}

\author{Gregor Wergen}
\email[]{gw@thp.uni-Koeln.de}
\affiliation{Institut f\"ur Theoretische Physik, Universit\"at zu
K\"oln, 50937 K\"oln, Germany}

\author{Satya N. Majumdar}
\email[]{satya.majumdar@u-psud.fr}
\affiliation{Laboratoire de Physique Th\'eorique et Mod\`eles Statistiques, UMR 8626, Universit\'e Paris Sud 11 and CNRS, B?t. 100, Orsay F-91405, France}

\author{Gr\'egory Schehr}
\email[]{gregory.schehr@u-psud.fr}
\affiliation{Laboratoire de Physique Th\'eorique et Mod\`eles Statistiques, UMR 8626, Universit\'e Paris Sud 11 and CNRS, B?t. 100, Orsay F-91405, France}

\noindent
\begin{abstract}

We study the statistics of the number of records $R_{n,N}$ for $N$ identical and 
independent symmetric discrete-time random walks of $n$ steps in one dimension, all 
starting at the origin at step $0$. At each time step,
each walker jumps by a random length drawn independently from a 
symmetric
and continuous distribution. We consider two cases: (I) when the variance $\sigma^2$
of the jump distribution is finite and (II) when $\sigma^2$ is divergent as in
the case of L\'evy flights with index $0 < \mu
< 2$.  
In both cases we find that the mean record number $\langle R_{n,N} \rangle$ grows 
universally as $\sim \alpha_N \sqrt{n}$ for large $n$, but with a very different behavior 
of the amplitude $\alpha_N$ for $N > 1$ in the two cases.
We find that for large $N$, $\alpha_N \approx 2 
\sqrt{\log N} $ independently of $\sigma^2$ in case I. In contrast, in case II,
the amplitude approaches to an $N$-independent constant for large $N$,
$\alpha_N \approx 4/\sqrt{\pi}$, independently of $0<\mu<2$.
For finite 
$\sigma^2$ we argue, and this is confirmed by our numerical simulations, that the 
full distribution of $(R_{n,N}/\sqrt{n} - 2 \sqrt{\log N}) \sqrt{\log N}$ converges to a 
Gumbel law as $n \to \infty$ and $N \to \infty$. In case II, our numerical simulations 
indicate that the distribution of $R_{n,N}/\sqrt{n}$ converges, for $n \to \infty$ and $N 
\to \infty$, to a universal nontrivial distribution, independently of $\mu$. We discuss 
the applications of our results to the study of the record statistics of 366 daily stock 
prices from the Standard \& Poors 500 index.

\end{abstract}

\maketitle
\section{Introduction}

A record is an entry in a series of events that exceeds all previous entries. In recent 
years there has been a surge of interest in the statistics of record-breaking events, 
both from the theoretical point of view as well as in multiple  
applications. The 
occurrence of record-breaking events has been studied for instance in sports 
\cite{Gembris2002,Gembris2007}, in evolution models in biology \cite{KJ,Krug1}, in 
the theory of spin-glasses \cite{Oliveira2005,Sibani2006} and in models of growing
networks\cite{GL1}. Recently there has been some progress in understanding the 
phenomenon of global warming via studying the occurrence of record-breaking temperatures
\cite{Redner2006,Meehl2009,WK,AK}.

More precisely, let us consider a sequence or a discrete-time series of random variables 
$\{x(0),x(1),x(2),\ldots, x(n)\}$ with $n+1$ entries. This sequence may represent for 
example the daily maximum temperature in a city or the daily maximum price of a stock. A 
record is said to happen at step $m$ if the $m$-th member of the sequence is bigger than 
all previous members, i.e., if $x(m)>x(i)$ for all $i=0,1,2,\ldots, (m-1)$. Let $R_n$ 
denote the number of records in this sequence of $n+1$ entries. Clearly $R_n$ is a random 
variable whose statistics depends on the joint distribution of $P(x(0),x(1),\ldots, 
x(n))$ of the members of the sequence. When the members of the sequence are independent 
and identically distributed (i.i.d) random variables each drawn from a distribution 
$p(x)$, i.e., the joint distribution factorizes, $P(x(0),x(1),\ldots, 
x(n))=\prod_{i=1}^{n+1} p(x(i))$, the record statistics is well understood from classical 
theories~\cite{FS54,Arnold,Nevzorov}. In particular, when $p(x)$ is a continuous distribution, it is known that the distribution of record number $P(R_n,n)$ is universal for all $n$, i.e., independent of the parent distribution $p(x)$. The average number of records up to step $n$, $\langle R_n\rangle = \sum_{m=1}^{n+1} 1/m$ for all $n$ and the universal distribution, for large $n$, converges to a Gaussian distribution with mean  $\approx \ln(n)$ and variance $\approx \ln n$. 

While the statistical properties of records for i.i.d 
random variables (RV's) are thus well understood for many years, 
numerous questions remain open for more realistic systems with time-dependent 
or correlated RV's. In principle there are many different ways to generalize the simple 
i.i.d. RV scenario described above. For instance, one can consider time series of RV's 
that are independent, but not identically distributed. One example for this case is the 
so called Linear Drift Model with RV's from probability distributions with identical 
shape, 
but with a mean value that increases in time. This model was first proposed in the 1980's 
\cite{Ballerini1985} and was recently thoroughly analyzed in Refs. 
\cite{FWK1, WFK, FWK2}. In 2007 Krug also considered 
the case of uncorrelated RV's from distributions 
with increasing variance \cite{Krug1}.

Another possible generalization is the one where RV's are correlated. Perhaps, the
simplest and the most natural
model of 
correlated RV's is an $n$-step one dimensional discrete-time random walk with
entries $\{x(0)=0,x(1),x(2),\ldots,x(n)\}$ where the position 
$x(m)$ of the walker at discrete time $m$ evolves 
via 
the Markov 
jump process
\begin{equation}
x(m)= x(m-1) + \eta(m) \;,
\label{markov.1}
\end{equation}
with $x(0)=0$ and $\eta(m)$ represents the random jump at step $m$. The
noise variables $\eta(m)$'s are assumed to be i.i.d variables, each 
drawn from a symmetric distribution $f(\eta)$. 
For instance, it may include L\'evy flights where $f(\eta)\sim 
|\eta|^{-1-\mu}$ for large $\eta$ with the L\'evy index $0<\mu< 2$
which has a divergent second moment. 
Even though this model represents a very simple Markov chain, statistical properties of 
certain observables
associated with such a walk may be quite nontrivial to compute, depending on 
which observable one is studying  
\cite{Weiss,Redner,SMreview}. For instance, in recent years there has been a lot of 
interest in the 
extremal properties of such random walks. These include the statistics of 
the maximal 
displacement of the walk up to $n$ steps with several 
applications~\cite{CFFH,CM2005,MCZ2006,SMreview,FM2012} and the order 
statistics, i.e., the statistics of the ordered maxima~\cite{MOR11,SM12} 
as well as the universal 
distribution of gaps
between successive ordered maxima of a random walk~\cite{SM12}.

The statistics of the number of 
record-breaking events in the discrete-time random 
walk process in Eq. (\ref{markov.1}) has also been studied 
in a number of recent works with 
several interesting results~\cite{MZrecord,PLDW2009,SS,WBK,EKB2011}. 
In 2008, Majumdar and Ziff computed exactly the full distribution
$P(R_n,n)$ of the record number up to $n$ steps and found that
when the jump distribution $f(\eta)$ is continuous and symmetric, the
record number distribution $P(R_n,n)$ is completely universal for all $n$,
i.e., independent of the details of the jump distribution~\cite{MZrecord}. 
In particular, for instance, the L\'evy flight with index $0<\mu<2$ 
(thus with a divergent second moment of the jump distribution $f(\eta)$)
has the same record 
number
distribution as for a Gaussian walk (with a finite second moment of $f(\eta)$).
This is a rather 
amazing result
and the deep reason for this universality is rooted~\cite{MZrecord} in 
the so called
Sparre Andersen theorem~\cite{SA}. In particular, for large $n$,
$P(R_n,n)\sim n^{-1/2} G(R_n/\sqrt{n})$ where the scaling function
$G(x)= e^{-x^2/4}/\sqrt{\pi}$ is universal~\cite{MZrecord}.
The mean number of records $\langle R_n\rangle\approx \sqrt{4n/\pi}$ for large 
$n$~\cite{MZrecord}. 
In contrast, this universal result does not hold for symmetric but 
discontinuous $f(\eta)$. For example,  
if $f(\eta)= 
\frac{1}{2}\delta(\eta-1)+\frac{1}{2}\delta(\eta+1)$, then $x_m$ represents
the position of a random walker at step $m$ on a $1$-d lattice with lattice 
spacing $1$.
In this case, the mean number of records still grows as $\sqrt{n}$ for large $n$  
but with a smaller prefactor, $\langle R_n\rangle \approx 
\sqrt{2n/\pi}$~\cite{MZrecord}.

These results were later generalized
to several interesting cases, for instance, to the record statistics of
one dimensional random walk in presence of an external 
drift~\cite{PLDW2009,WBK} and
one dimensional continuous-time random 
walk with a waiting-time distribution between successive jumps~\cite{SS}.
The record statistics of the distance traveled by a random walker in 
higher dimensions with and without drift has been studied numerically
in the context of contamination spread in porous medium~\cite{EKB2011}. 
In \cite{WBK}, it was also found that the record statistics of stock markets 
is very similar to the ones of biased random walks.

While in Refs. \cite{MZrecord,PLDW2009,SS,WBK,EKB2011} the record statistics of a 
single 
discrete-time random walker was studied, the purpose of this article is to generalize
these results to the case where one has $N$ independent one dimensional discrete-time 
random walks. In this $N$-walker process, a record happens at an instant when the 
maximum position of all the walkers at that instant exceeds all its previous values.  
We will see that despite the fact that the walkers are independent, the record 
statistics is rather rich, universal and nontrivial even in this relatively
simple model.

Let us first summarize our main results. 
We derive asymptotic results for the mean of the record number $\langle 
R_{n,N} \rangle$ up to a time $n$ and also discuss its full distribution. 
It turns out that for $N>1$, while the full universality with respect 
to the jump distribution found for $N=1$ case is no longer valid, there
still remains a vestige of universality of a different sort.  
In our analysis, it is important to distinguish two cases: case (I) where
the jump distribution $f(\eta)$ has a finite variance $\sigma^2=\int_{-\infty}^{\infty}
\eta^2\, f(\eta)\, d\eta$ and case (II) where $\sigma^2$ is divergent as
in the case of L\'evy flights with L\'evy index $0<\mu<2$.
In both cases, we find that the mean record number $\langle R_{n,N} \rangle$ grows
{\em universally} as $\sim \alpha_N \sqrt{n}$ for large $n$. However, the $N$ 
dependence of the prefactor
$\alpha_N$, in particular for large $N$, turns out to be rather different in the two 
cases
\begin{equation}\label{ampN}
\alpha_N\xrightarrow[N\to \infty]{} \left\{\begin{array}{rl}2
\sqrt{\log N}  & \textrm{in Case I}\quad (\textrm{independent of}\, \sigma^2)
    \\
    \vspace{1.5mm}\\
  4/\sqrt{\pi} & \textrm{in Case II}\quad (\textrm{independent of}\, \mu)
\end{array}\right.
\end{equation}
In addition, we also study the distribution of the record number $R_{n,N}$.
For finite
$\sigma^2$ we argue and confirm numerically that the distribution of the random variable
$(R_{n,N}/\sqrt{n} - 2 \sqrt{\log N}) \sqrt{\log N}$ converges
to the Gumbel law asymptotically for large $n$ and $N$ (see section II for details).
In contrast, in case II, we find numerically that the distribution of $R_{n,N}/\sqrt{n}$ 
converges, for large $n$
and $N$, to a nontrivial distribution independent of the value of $0<\mu<2$ (see
section II for details). 
We were however unable to compute this asymptotic distribution analytically and it
remains a challenging open problem. Finally, we discuss
the applications of our results to the study of the 
record statistics of 366 daily stock
prices from the Standard \& Poors 500 index ~\cite{SP500}. 
We analyze the evolution of the record number in subsets of $N$ stocks that were randomly
chosen from this index and compare the results to our analytical findings. While the     
strong correlations between the individual stocks seem to play an important effect in the
record statistics, the dependence of the record number on $N$ still seems to be the same
as in the case of the $N$ independent random walkers.

The rest of the paper is organized as follows. In section II, we define the
$N$-walker model precisely and summarize the main results obtained
in the paper. In section III,
we present the analytical 
calculation of the mean number of records for multiple random walkers, in both
cases where $\sigma^2$ is finite (case I) and $\sigma^2$ is infinite (case II).
Section IV is devoted to an analytic study of the distribution of the record
number in the case where $\sigma^2$ is finite.
In section V we present a
thorough numerical study of the record statistics of multiple random walks, and
in section VI we discuss the application of our results to the record statistics
of stock prices.
Finally, we conclude in section VII and present the technical details of some of the 
analytical computations
concerning the
computation of the mean number of records and the distribution of the record
number for lattice random
walks in the three Appendices A, B and C..

\section{Record Statistics for Multiple Random Walks: The model and the main 
results}

\begin{figure}
\includegraphics[width=0.8\textwidth]{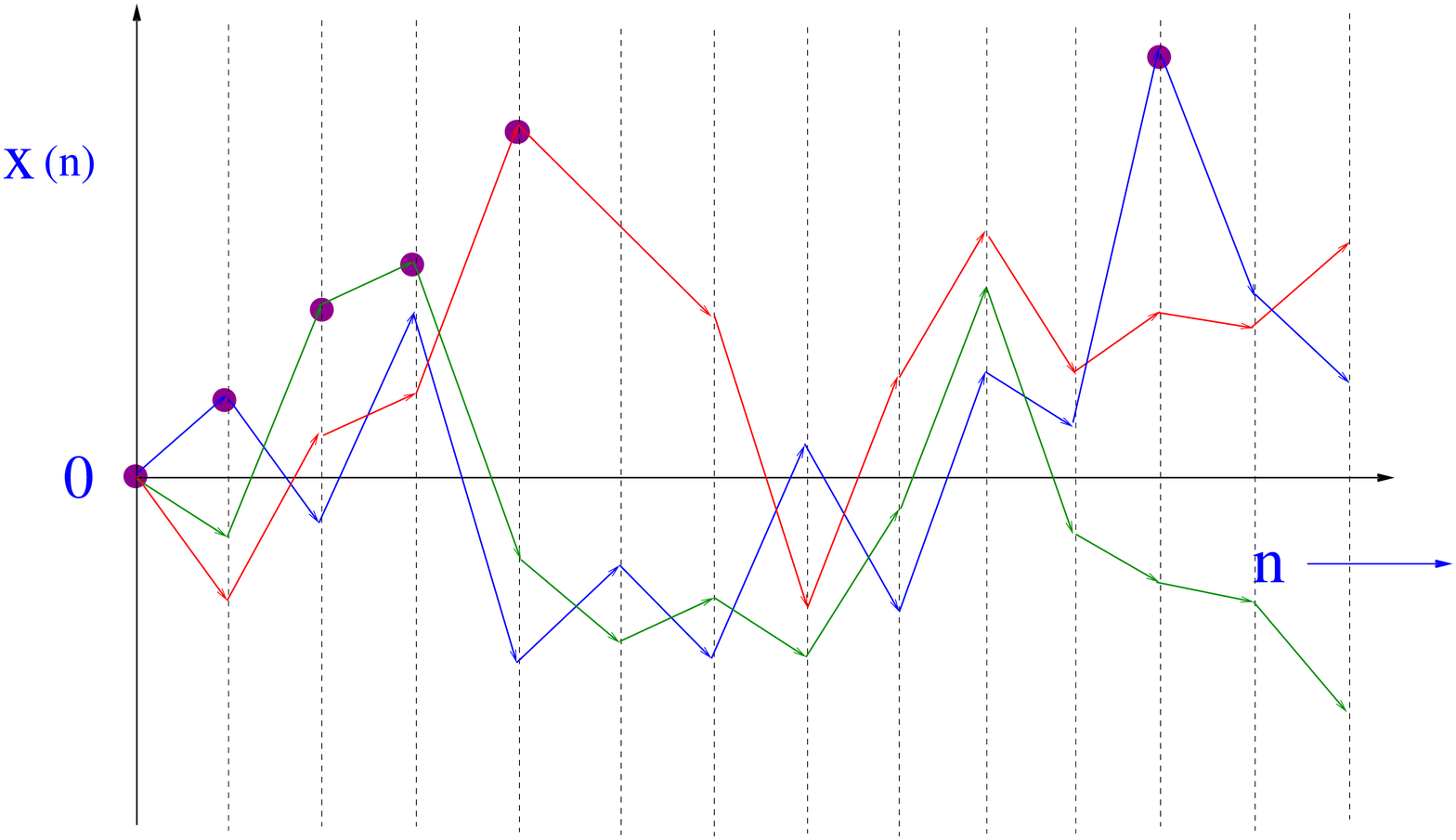}
\caption{Schematic trajectories of $N=3$ random walkers. Each walker starts
at the origin and evolves via the Markov jump process in Eq. (\ref{markov.2}).
A record happens at step $m$ if the maximum position at step $m$ 
$x_{\rm max}(m)> x_{\rm max}(k)$ for all $k=0,1,2,\ldots (m-1)$.
The record values are shown by filled circles.}
\label{multi1.fig} 
\end{figure}

Here we consider the statistics of records of $N$ independent random walkers all 
starting at the origin $0$. The position $x_i(m)$ of the $i$-th walker at 
discrete time step $m$ evolves via the Markov evolution rule
\begin{equation}
x_i(m)= x_{i}(m-1) +\eta_i(m) \;,
\label{markov.2}
\end{equation}
where $x_i(0)=0$ for all $i=1,2,\ldots, N$ and the noise $\eta_i(m)$'s are i.i.d
variables (independent from step to step and from walker to walker), each drawn
from a symmetric distribution $f(\eta)$. We are interested in the
record statistics of the composite process. More precisely, consider
at each step $m$, the maximum position of all $N$ random walkers
\begin{equation} 
x_{\rm max}(m)= {\rm max}\left[x_1(m), x_2(m),\ldots, x_N(m)\right].
\label{max.1}
\end{equation} 
A record is said to happen at step $m$ if this maximum position at step $m$
is bigger than all previous maximum positions, i.e. if
$x_{\rm max}(m)> x_{\rm max}(k)$ for all $k=0,1,\ldots, (m-1)$ (see Fig. 
\ref{multi1.fig}).
In other words, we are interested in the record statistics of
the stochastic discrete-time series $\{ x_{\rm max}(m)\}$, with the convention that the initial position $x_{\max}(0)=0$ is counted as a record. Note that even though
the position of each walker evolves via the simple independent Markovian rule in 
Eq. (\ref{markov.2}), the evolution of the maximum process $\{ x_{\rm max}(m)\}$ is
highly non-Markovian and hence is nontrivial.  

Let $R_{n,N}$ denote the number of records up to step $n$ for this
composite $N$-walker process. Clearly $R_{n,N}$ is a random variable
and we are interested in its statistics. For a single walker $N=1$,
we have already mentioned that the probability distribution
of the record number $R_{n,1}$ is completely universal, i.e.,
independent of the jump distribution $f(\eta)$ as long as
$f(\eta)$ is symmetric and continuous~\cite{MZrecord}. In particular,
for example, the
record number distribution is the same for simple Gaussian walkers
as well for L\'evy flights with index $0<\mu<2$. Here we are interested in the
opposite limit when $N\to \infty$.

We find that while the complete universality of the record statistics
is no longer true for $N>1$, a different type of universal behavior
emerges in the $N\to \infty$ limit. In this large $N$ limit,
there are two universal asymptotic behaviors of the record statistics depending
on whether the second moment $\sigma^2=\int_{-\infty}^{\infty} \eta^2\, 
f(\eta)\, d\eta $ of the jump distribution is finite or divergent.
For example, for Gaussian, exponential, uniform jump distributions
$\sigma^2$ is finite. In contrast, for L\'evy flights where $f(\eta)\sim 
|\eta|^{-\mu-1}$ for large $\eta$ with the L\'evy index $0<\mu<2$, the second
moment $\sigma^2$ is divergent. In these two cases, we find the
following behaviors for the record statistics.

\vskip 0.3cm

\noindent {\bf Case I ($\sigma^2$ finite):} In this case, we 
consider jump distributions $f(\eta)$ that are symmetric with
a finite second moment $\sigma^2=\int_{-\infty}^{\infty} \eta^2\, f(\eta)\, 
d\eta $. In this case, the Fourier transform of the jump distribution
${\hat f}(k)= \int_{-\infty}^{\infty} f(\eta)\,e^{ik\eta}\, d\eta$ behaves,
for small $k$, as
\begin{equation}
{\hat f}(k) \approx 1- \frac{\sigma^2}{2}\, k^2 + \ldots
\label{fourier1}
\end{equation} 
Examples include the Gaussian jump distribution, 
$f(\eta)=\sqrt{a/\pi}\,e^{-a\,\eta^2} $, exponential jump distribution
$f(\eta)= (b/2)\, \exp[-b|\eta|] $, uniform jump distribution over $[-l,l]$
etc. For such jump distributions, 
we find that
for large number of walkers $N$, the mean number of 
records grows asymptotically for large $n$ as
\begin{equation}
\langle R_{n,N}\rangle \xrightarrow[N\to \infty]{n\to \infty} 2\, \sqrt{\ln N}\, 
\sqrt{n} \;.
\label{meanrecord1}
\end{equation}
Note that this asymptotic behavior is universal in the sense that it does not 
depend explicitly on $\sigma$ as long as $\sigma$ is finite.

Moreover, we argue that for large $N$ and large $n$, the scaled random variable 
$R_{n,N}/\sqrt{n}$ converges, in distribution, to the Gumbel form, i.e,
\begin{equation}
{\rm Prob.}\left[ \frac{R_{n,N}}{\sqrt{n}}\le x\right] \xrightarrow[N\to 
\infty]{n\to 
\infty} F_{1}\left[\left(x-2\,\sqrt{\ln N}\right)\,\sqrt{\ln N}\right] \;, \quad 
{\rm 
where} \quad 
F_1(z)=\exp\left[-\exp[-z]\right].
\label{gumbel.1}
\end{equation}
Indeed, for large $N$ and large $n$, the scaled variable 
$R_{n,N}/\sqrt{n}$ converges, in distribution, to the maximum of $N$
independent random variables
\begin{equation}
\frac{R_{n,N}}{\sqrt{n}}\xrightarrow[N\to \infty]{n\to
\infty} M_N\, \quad {\rm where}\quad M_N= {\rm max}(y_1,y_2,\ldots, y_N) 
\label{gumbel.2}
\end{equation}
where $y_i\ge 0$'s are i.i.d non-negative random variables each drawn from  
distribution $p(y)= \frac{1}{\sqrt{\pi}}\, e^{-y^2/4}$ for $y\ge 0$ and 
$p(y)=0$ for $y<0$.

\vskip 0.3cm

\noindent {\bf Case II ($\sigma^2$ divergent ):} In this case we consider jump 
distributions $f(\eta)$ such that the second moment $\sigma^2$ is divergent.
In this case, the Fourier transform ${\hat f}(k)$ of the noise distribution 
behaves, for all 
$k$, as
\begin{equation}
{\hat f}(k) = 1- |a\,k|^{\mu} + \ldots 
\label{fourier2}
\end{equation}
where $0<\mu<2$. 
Examples include L\'evy flights where $f(\eta)\sim |\eta|^{-\mu-1}$ with the
L\'evy index $0<\mu< 2$. For the noise distribution in Eq. (\ref{fourier2}), we 
find, quite amazingly, that
in the large $N$ and large $n$ limit, the record statistics is (i) completely 
universal, i.e., independent of $\mu$ and $a$ 
(ii) more surprisingly and unlike in Case-I, the record statistics also
becomes independent of $N$ as $N\to \infty$. For example, we prove that
for large $N$, the mean number of records grows asymptotically with $n$
as
\begin{equation}
\langle R_{n,N}\rangle \xrightarrow[N\to \infty]{n\to \infty} 
\frac{4}{\sqrt{\pi}}\, \sqrt{n} \;,
\label{meanrecord2}
\end{equation} 
which is exactly twice that of one walker, i.e., $\langle R_{n,N\to 
\infty}\rangle = 2\, \langle R_{n,1}\rangle$ for large $n$.
Similarly, we find that the scaled variable $R_{n,N}/\sqrt{n}$, for large $n$
and large $N$, converges to a universal distribution
\begin{equation}
{\rm Prob.}\left[ \frac{R_{n,N}}{\sqrt{n}}\le x\right] \xrightarrow[N\to
\infty]{n\to
\infty} F_2(x) \;,
\label{distri2}
\end{equation}
which is independent of the L\'evy index $\mu$ as well as of the scale $a$
in Eq. (\ref{fourier2}). 
While we have computed this universal distribution $F_2(x)$ numerically rather 
accurately,
we were not able to compute its analytical form.

\section{Mean Number of Records for Multiple Walkers}

Let $R_{n,N}$ be the number of records up to step $n$ for $N$ random walkers, 
i.e., for the maximum process $x_{\rm max}(n)$. Let us write
\begin{equation}
R_{m,N}= R_{m-1,N}+ \xi_{m,N} \;,
\label{recur.1}
\end{equation}
where $\xi_{m,N}$ is a binary random variable taking values $0$ or $1$.
The variable $\xi_{m,N}=1$ if a record happens at step $m$ and 
$\xi_{m,N}=0$
otherwise. Clearly, the total number of records up to step $n$ is
\begin{equation}
R_{n,N}= \sum_{m=1}^n \xi_{m,N} \, .
\label{recordnumber.1}
\end{equation} 
So, the mean number of records up to step $n$ is
\begin{equation}
\langle R_{n,N} \rangle = \sum_{m=1}^n \langle \xi_{m,N}\rangle= \sum_{m=1}^n 
r_{m,N} \;,
\label{mean.1}
\end{equation}
where $r_{m,N}= \langle \xi_{m,N}\rangle$ is just the record rate, i.e., the 
probability that a record happens at step $m$. To compute the mean number
of records, we will first evaluate
the record rate $r_{m,N}$ and then sum over $m$.

\begin{figure}
\includegraphics[width=0.8\textwidth]{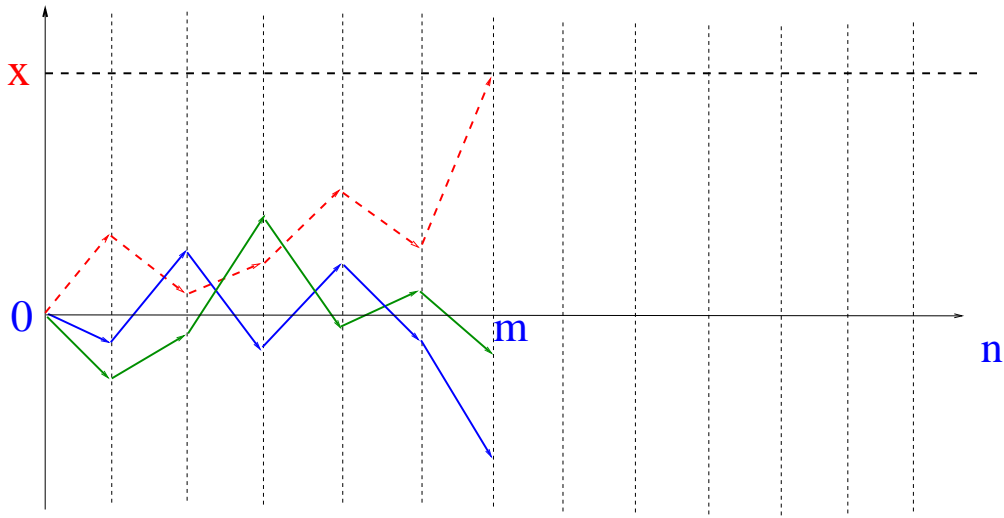}
\caption{A record happens at step $m$ with record value $x$ for $N=3$ walkers, 
all starting at the origin. 
This event corresponds to one walker (the dashed line) reaching the level $x$
for the first time at step $m$ while the other walkers stay below the level
$x$ up to step $m$.}
\label{recrate1.fig} 
\end{figure}

To compute $r_{m,N}$ at step $m$, we need to sum 
the probabilities of all trajectories that lead to a record event at step $m$.
Suppose that a record happens at step $m$ with the record value $x$ (see Fig. 
\ref{recrate1.fig}). This  corresponds to the event that one of the $N$ walkers 
(say the dashed trajectory
in Fig. \ref{recrate1.fig}), starting at the origin at step $0$, has reached 
the level $x$ for the first time
at step $m$, while the rest of the $N-1$ walkers, starting at the origin at step 
$0$, have all stayed below the level $x$ till the step $m$. Also, the walker
that actually reaches $x$ at step $m$ can be any of the $N$ walkers.
Finally this event can take place at any level $x>0$ and one needs to
integrate over the record value $x$. 
Using the independence of $N$ walkers and taking into account the event
detailed above, one can 
then write 
\begin{equation}
r_{m,N} = N\, \int_0^{\infty} p_m(x)\, \left[q_m(x)\right]^{N-1}\, dx \;,
\label{rate.1}
\end{equation}
where $q_m(x)$ denotes the probability that a single walker, starting at the
origin, stays below the level $x$ up to step $m$ and $p_m(x)$ is the
probability density that a single walker reaches the level $x$ for the 
first time at step $m$, starting at the origin at step $0$. 

The two quantities $p_m(x)$ and $q_m(x)$ can be reinterpreted in terms
of slightly more
familiar objects via the following observation.
Note that by shifting the origin to the level $x$
and using the time-reversal property of the trajectory of a single random walker,
it is easy to see that $p_m(x)$ is just the probability density that a single
walker, starting at the origin at step $0$, reaches $x$ at step $m$ while
staying positive at all intermediate steps.
By a similar shift of the origin to level $x$ and using the reflection symmetry 
of the trajectories around the origin,
it is clear that $q_m(x)$ can be interpreted as the probability
that a single walker, starting at an initial position $x>0$ at step $0$,
stays positive (i.e., does not cross the origin) up to step $m$.
This is then the familiar persistence or the survival probability of a single 
random
walker~\cite{SMreview}. 
In fact, both these quantities $p_m(x)$ and $q_m(x)$ can be regarded as
special cases of the more general restricted Green's function in the
following sense.
Consider a single random walker starting at position $x$ at step $0$
and evolving its position via successive uncorrelated jumps as 
in Eq. (\ref{markov.1}). Let
$G_{+}(y,x,m)$ denote the probability density that the walker
reaches $y>0$ at step $m$, starting at $x>0$ at step $0$, while staying positive 
at all intermediate
steps. The subscript $+$ denotes that it is indeed the restricted Green's
function counting only the trajectories that reaches $y$ at step $m$
without crossing the origin in between. It is then clear from
our discussion above that
\begin{eqnarray}
p_m(x) &=& G_+(x,0,m)  \label{pm1} \\
q_m(x)& = & \int_0^{\infty} G_+(y,x,m)\, dy \,. \label{qm1} 
\end{eqnarray} 
In the second line, the survival probability $q_m(x)$ is obtained from the 
restricted Green's function by 
integrating over all possible final positions of the walker.
Note also, from Eqs. (\ref{pm1}) and (\ref{qm1}), that the survival
probability starting exactly at the origin is
\begin{equation}
q_m(0)= \int_0^{\infty} p_m(x)\,dx.
\label{q0}
\end{equation}

Hence, if we know the restricted Green's function $G_+(y,x,m)$, we can in 
principle compute the two required quantities $p_m(x)$ and $q_m(x)$.
Using the Markov evolution rule in Eq. (\ref{markov.1}), it is easy to
see that
the restricted Green's function $G_+(y,x,m)$ satisfies an integral
equation in the semi-infinite domain~\cite{SMreview} 
\begin{equation}
G_+(y,x,m)= \int_0^{\infty} G_+(y', x,m-1)\, f(y-y')\, dy' \;,
\label{WH1}
\end{equation}
starting from the initial condition, $G_+(y,x,0)= \delta(y-x)$.
Such integral equations over the semi-infinite domain are called
Wiener-Hopf equations and are notoriously difficult to solve for arbitrary kernel 
$f(z)$. Fortunately, for the case when $f(z)$ represents
a continuous and symmetric probability density as in our case, one can obtain 
a closed form solution for the following generating function (rather its Laplace 
transform)~\cite{Ivanov}
\begin{equation}
\int_0^{\infty}dy\, e^{-\lambda\,y} \int_0^{\infty} dx\, e^{-\lambda_0\, 
x}\, \left[\sum_{m=0}^{\infty} G_+(y,x,m)\, s^m\right]=
{\tilde G}(\lambda,\lambda_0,s)= 
\frac{\phi(s,\lambda)\,\phi(s,\lambda_0)}{\lambda+\lambda_0}\;,
\label{ivanov.1}
\end{equation}
where 
\begin{equation}
\phi(s,\lambda)= \exp\left[-\frac{\lambda}{\pi}\,\int_0^{\infty} \frac{\ln [1-s 
{\hat f}(k)]}{\lambda^2+k^2}\, dk\right] \quad {\rm and}\quad {\hat f}(k)= 
\int_{-\infty}^{\infty} f(x)\, e^{i\,k\,x}\, dx \;.
\label{ivanov.2}
\end{equation}

While the formula in Eq. (\ref{ivanov.1}) is explicit, it is rather cumbersome
and one needs further work to extract the asymptotic behavior of $p_m(x)$
and $q_m(x)$ from this general expression. To make progress, one can first 
make a change of variable on the left hand side (lhs) $\lambda_0 x=z$ and
then take the $\lambda_0\to \infty$ limit. Using $\phi(s,\lambda_0\to \infty)=1$
and the definition $G_+(y,0,m)=p_m(y)$, and replacing $y$ by $x$ we then obtain 
the following
relation
\begin{equation}
\sum_{m=0}^{\infty}s^m\, \int_0^{\infty} p_m(x)\, e^{-\lambda x}\, dx= 
\phi(s,\lambda)
\label{pm.2}
\end{equation}
where $\phi(s,\lambda)$ is given in Eq. (\ref{ivanov.2}). Similarly, putting
$\lambda=0$ on the lhs of Eq. (\ref{ivanov.1}), using the definition
$q_m(x)= \int_0^{\infty} G_+(y,x,m)\,dy$ and replacing $\lambda_0$ by
$\lambda$, it is easy 
to see that
\begin{equation}
\sum_{m=0}^{\infty} s^m\, \int_0^{\infty} q_m(x)\, e^{-\lambda x}\, dx= 
\frac{1}{\lambda \sqrt{1-s}}\, \phi(s,\lambda)\, .
\label{qm.2}
\end{equation}
The formula in Eq. (\ref{qm.2}) is known in the literature as the 
celebrated Pollaczek-Spitzer formula~\cite{Pollaczek,Spitzer} and has 
been used in a number of works to derive exact results on the maximum of a 
random jump process~\cite{Darling,CM2005,KMR2011,GRS2011}. Interestingly,
this formula has also been useful to compute 
the asymptotic behavior of
the flux of particles to a spherical trap in three 
dimensions~\cite{MCZ2006,ZMC2007,ZMC2009}. 

Let us also remark that by making a change of variable $\lambda x=y$ on 
the lhs of Eq. (\ref{qm.2}) and taking $\lambda 
\to \infty$, one obtains the rather amazing universal result for all $m$
\begin{equation}
\sum_{m=0}^{\infty} q_m(0)\, s^m = \frac{1}{\sqrt{1-s}} \Longrightarrow
q_m(0)= {{2m}\choose m}\frac{1}{2^{2m}} \;,
\label{sa1}
\end{equation}
which is known as the Sparre Andersen theorem~\cite{SA}. In particular, for large 
$m$, $q_m(0)\approx 1/\sqrt{\pi m}$.
note that for the case of a single walker $N=1$, it follows from
Eq. (\ref{rate.1}) that the record 
rate at step $m$ is simply given by
\begin{equation}
r_{m,1}= \int_0^{\infty}p_m(x)\, dx= q_m(0)= {{2m}\choose m}\frac{1}{2^{2m}}
\xrightarrow{m\to \infty} \frac{1}{\sqrt{\pi m}} \;,
\label{N1.1}
\end{equation}
where we have used Eq. (\ref{q0}) and the Sparre Andersen theorem (\ref{sa1}).
Thus, one obtains the rather surprising universal result for the $N=1$ case:
for all continuous and symmetric jump distributions, the mean number
of records up to step $n$, $\langle R_{n,N}\rangle = \sum_{m=1}^n r_{m,N}$ is 
universal
for all $n$ and grows as $\sqrt{4n/\pi}$ for large $n$~\cite{MZrecord}.
The universality in this case can thus be traced back to Sparre Andersen
theorem.

In contrast, for $N>1$, we need the full functions $p_m(x)$ and $q_m(x)$
to compute the record rate in Eq. (\ref{rate.1}). This is hard to compute 
explicitly for all $m$. However, one can make progress in computing the 
asymptotic behavior of the record rate
$r_{m,N}$ for large $m$ and large $N$, as we show below.   
In turns out that for large $m$, the integral in Eq. (\ref{rate.1}) is dominated
by the asymptotic scaling behavior of the two functions $p_m(x)$ and $q_m(x)$
for large $m$ and large $x$.
To extract the scaling behavior of $p_m(x)$ and $q_m(x)$,
our starting point would be the two equations 
(\ref{pm.2}) and (\ref{qm.2}).
The next step is to use these asymptotic expressions in the main formula
in Eq. (\ref{rate.1}) to determine the record rate $r_{m,N}$ at step $m$ for
large $m$ and large $N$. The procedure to extract the asymptotics is 
somewhat subtle and algebraically cumbersome. To facilitate an easy reading 
of the paper, we relegate this algebraic procedure in the appendices.
Here we just use the main results from these appendices and proceed to 
derive the results announced in Eqs. (\ref{meanrecord1}) and (\ref{meanrecord2}).
The asymptotic behavior of $p_m(x)$ and $q_m(x)$ depend on whether
$\sigma^2=\int_{-\infty}^{\infty} \eta^2 \, f(\eta)\, d\eta$ is finite or 
divergent. This gives rise to the two cases mentioned in Section II.

\vskip 0.3cm

\noindent {\bf Case I($\sigma^2$ finite):} In this case, we show in Appendix 
A that in the scaling limit $x\to \infty$, $m\to \infty$ but keeping the
ration $x/\sqrt{m}$ fixed, $p_m(x)$ and $q_m(x)$ approach the following 
scaling behavior
\begin{eqnarray}
p_m(x)&\to & \frac{1}{\sqrt{2\sigma^2}\, m}\, 
g_1\left(\frac{x}{\sqrt{2\,\sigma^2\, 
m}}\right)\,,
\quad 
{\rm where}\quad g_1(z)= \frac{2}{\sqrt{\pi}}\,z\, e^{-z^2} \;, \label{pm1scaling} \\
q_m(x) & \to & h_1\left(\frac{x}{\sqrt{2\,\sigma^2\,
m}}\right)\,,
\quad
{\rm where}\quad h_1(z)= {\rm erf}(z) \;,\label{qm1scaling}
\end{eqnarray}
where ${\rm erf}(z)= \frac{2}{\sqrt{\pi}}\, \int_0^z e^{-u^2}\, du$. Note that
$dh_1(z)/dz= g_1(z)/z$.

\vskip 0.3cm

\noindent {\bf Case II ($\sigma^2$ divergent):} For the case when the 
Fourier
transform of the jump distribution ${\hat f}(k)$ has the small $k$ behavior
as in Eq. (\ref{fourier2}), we show in Appendix B that in the scaling limit
when $x\to \infty$, $m\to \infty$, but keeping the ratio $x/m^{1/\mu}$ fixed,
\begin{eqnarray}
p_m(x)&\to & \frac{1}{m^{1/2+1/\mu}}\, g_2\left(\frac{x}{m^{1/\mu}}\right) \label
{pm2scaling} \\
q_m(x) & \to & h_2\left(\frac{x}{m^{1/\mu}}\right). \label
{qm2scaling} 
\end{eqnarray}
While it is hard to obtain explicitly the full scaling functions $g_2(z)$ 
and $h_2(z)$ for all $z$, one can compute the large $z$ asymptotic
behavior and obtain
\begin{eqnarray}
g_2(z) &\underset{z \to \infty}{\sim} & \frac{A_{\mu}}{z^{1+\mu}} \;,\label{g2largez} \\
h_2(z) & \underset{z \to \infty}{\sim}  & 1- \frac{B_{\mu}}{z^{\mu}} \label{h2largez}
\end{eqnarray}
where the two amplitudes are 
\begin{eqnarray}
A_{\mu} &= & \frac{2\mu}{\sqrt{\pi}}\, \beta_{\mu} \label{amu1} \\
B_{\mu} &=& \beta_{\mu} \label{bmu1}
\end{eqnarray}
with the constant $\beta_{\mu}$ having different expressions for
$0<\mu<1$ and $1\le \mu <2$
\begin{eqnarray}
\beta_{\mu}&= & \frac{a^{\mu}}{\pi 
\Gamma(1-\mu)}\,\int_0^{\infty}\frac{u^{\mu}}{1+u^2}\, du \, \quad{\rm for}\quad 
0<\mu<1 \label{betamu1} \\
\beta_{\mu}& =& \frac{2a^\mu}{\pi \Gamma(2-\mu)}\,\int_0^{\infty} 
\frac{u^{\mu}}{(1+u^2)^2}\, du \, \quad{\rm for}\quad
1\le \mu<2 \label{betamu2} \, . 
\end{eqnarray} 
The expressions above (\ref{betamu1}, \ref{betamu2}) can be written in a unified way for any $0 < \mu < 2$ as
\begin{eqnarray}\label{unif_beta}
\beta_{\mu} = \frac{a^\mu}{2 \Gamma(1-\mu) \cos(\frac{\mu \pi}{2})} = \frac{a^\mu \Gamma(\mu) \sin{(\frac{\mu \pi}{2})}}{\pi} \;,
\end{eqnarray}
where, in the last equality, we have used $\Gamma(1-\mu) \Gamma(\mu) = \dfrac{\pi}{\sin{\mu \pi}}$. We recall that here we are considering discrete time random walks (\ref{markov.1}). In the continuous time random walk framework, with an exponential waiting time between jumps, the quantity $g_2(z)$ was studied in Ref. \cite{ZK95}. By performing an asymptotic analysis similar to the one presented in Appendix B, the authors showed that $g_2(z)$ behaves, for large $z$, like in Eq. (\ref{g2largez}) with the same exponent albeit with a different amplitude. On the other hand, the exact asymptotic result (\ref{g2largez}), together with~Eq. (\ref{unif_beta}) can also be used to study the normalized pdf $\tilde p_m(x)$ of the position after $m$ steps, with the condition that the walker stays positive at all intermediate steps, which was recently studied in Ref. \cite{GRS2011}. It reads
\begin{eqnarray}
\tilde p_m(x) = \frac{p_m(x)}{\int_0^\infty p_m(x) dx}\to \frac{1}{m^{1/\mu}} \tilde g_2\left(\frac{x}{m^{1\mu}}\right) \;, \; \tilde g_2(z) = \sqrt{\pi} g_2(z) \;, 
\end{eqnarray}
where we have used the Sparre-Andersen theorem $\int_0^\infty p_m(x) \, dx = q_m(0) \sim 1/\sqrt{\pi m}$ for large $m$. From Eq. (\ref{g2largez}), one obtains the large $z$ behavior of $\tilde g_2(z)$ as
\begin{eqnarray}\label{asympt_constrained_propag}
\tilde g_2(z) \underset{z \to \infty}{\sim} \frac{\tilde A_{\mu}}{z^{1+\mu}} \;, \; \tilde A_{\mu} =  \frac{2 a^\mu \sin\left( \frac{\mu \pi}{2}\right) \Gamma(\mu+1)}{\pi} \;,
\end{eqnarray}
where we have used $\mu \Gamma(\mu) = \Gamma(\mu+1)$. On the other hand, if one considers  the probability density function $P_{m}(x)$ of the position of a free L\'evy random walk after $m$ steps, with a jump distribution as in Eq.~(\ref{fourier2}) after $m$ steps, it assumes the scaling form, valid for large $m$, $P_m(x) \sim m^{-1/\mu} p(x/m^{1/\mu})$ where the asymptotic behavior is given by
\begin{eqnarray}\label{asympt_free}
p(z) \underset{z \to \infty}{\sim} \frac{C_\mu}{z^{1+\mu}} \;, \; C_\mu = \frac{a^\mu \sin\left( \frac{\mu \pi}{2}\right) \Gamma(\mu+1)}{\pi} \;.
\end{eqnarray}
Therefore the above result (\ref{asympt_constrained_propag}) establishes that $\tilde A_\mu = 2 C_\mu$: this result was recently obtained analytically in perturbation theory for $\mu$ close to $2$, $2 - \mu \ll 1$, and conjectured to hold for any $\mu$, on the basis of thorough numerical simulations \cite{GRS2011}. Here this result is established exactly for any $\mu \in (0,2)$. While the large $z$ behavior of $g_2(z)$ is the most relevant one for our study, we mention, for completeness, that its small $z$ behavior was also studied in Ref. \cite{ZK95,ZRM2009}, yielding $g_2(z) \sim z^{\mu/2}$. Finally we remark that the asymptotic behavior of $h_2(z)$ for large $z$
has been computed in great detail recently in Ref. \cite{KMR2011}, only the first two leading terms are presented in Eq. (\ref{h2largez}) here. 

We are now ready to use these asymptotic behavior of $p_m(x)$ and $q_m(x)$
in Eq. (\ref{rate.1}) to deduce the large $m$ behavior of the record rate.
Noting that for large $m$, the integral is dominated by the scaling regime,
we substitute in Eq. (\ref{rate.1}) the scaling forms of $p_m(x)$ and $q_m(x)$ 
found in Eqs. (\ref{pm1scaling}), (\ref{qm1scaling}), (\ref{pm2scaling}) and 
(\ref{qm2scaling}). We then get, for large $m$,
\begin{equation}
r_{m,N}\approx  \frac{N}{\sqrt{m}}\, \int_0^{\infty} g(z)\, 
[h(z)]^{N-1}\, dz \;,
\label{rate.2}
\end{equation}
where $g(z)=g_{1,2}(z)$ and $h(z)=h_{1,2}(z)$ depending on the two cases.
So, we notice that in all cases the record rate decreases as
$m^{-1/2}$ for large $m$, albeit with different $N$-dependent prefactors in  
the two cases. Hence, the mean number of records $\langle R_{n,N}\rangle $ up to 
step $n$
grows, for large $n$, as
\begin{equation}
\langle R_{n,N}\rangle  \approx \alpha_N\, \sqrt{n} \,, \quad {\rm where}\quad 
\alpha_N= 2 N\, 
\int_0^{\infty} g(z)\, [h(z)]^{N-1}\, dz \;.
\label{avgrec.1}
\end{equation}

Next we estimate the constant $\alpha_N$ for large $N$. We first note
that $\alpha_N$ in Eq. (\ref{avgrec.1}) can be expressed as
\begin{equation}
\alpha_N= 2\, \int_0^{\infty} \frac{g(z)}{h'(z)}\, \frac{d}{dz} \{[h(z)]^N\}\, dz \;,
\label{alphaN}
\end{equation}
where $h'(z)= dh/dz$. Noticing that $h(z)$ is an increasing function of 
$z$ approaching $1$ as $z\to \infty$, the dominant contribution to the 
integral for large $N$ comes from the large $z$ regime. Hence, we need
to estimate how the ratio $g(z)/h'(z)$ behaves for large $z$. Let us 
consider the two cases separately.

\vskip 0.3cm

\noindent {\bf Case I ($\sigma^2$ finite): } In this case, we have
explicit expressions for $g_1(z)$ and $h_1(z)$ respectively in Eqs. 
(\ref{pm1scaling}) and (\ref{qm1scaling}). Hence we get
\begin{eqnarray}
\alpha_N &=& 2\, \int_0^{\infty} dz\, z\, \frac{d}{dz} [{\rm erf}(z)]^N  \\
&=& \int_0^{\infty} dy\, y\, \frac{d}{dy} [{\rm erf}(y/2)]^{N}.
\label{alphaN1.case1}
\end{eqnarray}
The rhs of Eq. (\ref{alphaN1.case1}) has a nice interpretation.
Consider $N$ i.i.d positive random variables $\{y_1,y_2,\ldots, y_N\}$, each 
drawn from the
distribution: $p(y)= \frac{1}{\sqrt{\pi}}\,e^{-y^2/4}$ for $y\ge 0$
and $p(y)=0$ for $y<0$. Let $M_N$ 
denote their 
maximum. Then the cdf of the maximum is given by
\begin{equation}
{\rm Prob}[M_N\le y]= \left[\int_0^{y} p(y')\,dy'\right]^N= [{{\rm erf}(y/2}]^N\,.
\label{maximum.1}
\end{equation}  
The probability density of the maximum is then given by: $\frac{d}{dy} [{{\rm 
erf}(y/2}]^N$. Hence, the rhs of Eq. (\ref{alphaN1.case1}) is just the average
value $\langle M_N\rangle $ of the maximum. This gives us an identity 
for all $N$
\begin{equation}
\alpha_N= \langle M_N\rangle \;.
\label{maximum.2}
\end{equation}
From the standard extreme value analysis of i.i.d variables \cite{gumbel}, it is easy to show
that to leading order for large $N$, $\langle M_N\rangle \approx 2\sqrt{\ln N}$
which then gives, via Eq. (\ref{avgrec.1}), the leading asymptotic 
behavior of the mean record number
\begin{equation}
\langle R_{n,N} \rangle \xrightarrow[N\to \infty]{n\to \infty} 2 \sqrt{\ln N}\, \sqrt{n} \;.
\label{avgrec.case1}
\end{equation} 

\vskip 0.3cm

\noindent {\bf Case II ($\sigma^2$ divergent): } To evaluate $\alpha_N$ 
in Eq. (\ref{alphaN}), we note that when $\sigma^2$ is divergent, unlike in Case
I,
we do not have the full explicit form of the scaling functions $g_2(z)$
and $h_2(z)$. Hence evaluation of $\alpha_N$ for all $N$ is difficult.
However, we can make progress for large $N$. As mentioned before, for large $N$,
the dominant contribution to the integral in Eq. (\ref{alphaN}) comes
from large $z$. For large $z$, using the asymptotic expressions in Eqs. 
(\ref{g2largez}) and (\ref{h2largez}), we get
\begin{equation}
\frac{g_2(z)}{h_2'(z)}\xrightarrow{z\to \infty} \frac{A_\mu}{\mu\, 
B_{\mu}}=\frac{2}{\sqrt{\pi}} \;,
\label{ratiolargez}
\end{equation}
where we have used Eqs. (\ref{amu1}) and (\ref{bmu1}) for the expressions
of $A_\mu$ and $B_{\mu}$. We next substitute this asymptotic constant
for the ratio $g_2(z)/h_2'(z)$ in the integral on the rhs of Eq. (\ref{alphaN}).
The integral can then be performed trivially and we get, for large $N$,
\begin{equation}
\alpha_N \xrightarrow{N\to \infty} \frac{4}{\sqrt{\pi}}.
\label{alphaN.case2}
\end{equation}
From Eq. (\ref{avgrec.1}) we then get for the mean record number
\begin{equation}
\langle R_{n,N} \rangle \xrightarrow[N\to \infty]{n\to \infty} 
\frac{4}{\sqrt{\pi}}\, \sqrt{n} \, .
\label{avgrec.case2}
\end{equation} 
In contrast to case I in Eq. (\ref{avgrec.case1}), here 
the mean record number becomes independent of $N$ for large $N$. 

\section{The distribution of the number of records for finite $\sigma^2$}

In the previous section, we performed a very precise study of the mean number of 
records $\langle R_{n,N}\rangle$ up to step $n$, in both cases where $\sigma^2$ is finite 
and 
divergent. In the 
present section, we investigate the full probability distribution function (pdf) of the 
record number $R_{n,N}$. However, we have been able to make analytical
progress for the record number distribution only in case I where $\sigma^2$ is finite
to which we restrict ourselves below.

The clue that leads to an analytical computation of the record number distribution
is actually already contained in the exact expression of the mean record number
in Eqs. (\ref{avgrec.1}) and 
(\ref{alphaN1.case1}).
This result suggests that there perhaps is a relation between the record number
$R_{n,N}$ and the stochastic variable $Y_{n,N}$ defined as
\begin{equation}
Y_{n,N}= \max_{0 \leq m \leq n} x_{\max}(m) = \max_{0 \leq m \leq n} \max_{0 \leq i \leq 
N} [x_i(m)].
\label{ydef}
\end{equation}
Note that $Y_{n,N}$ simply denotes the maximum position of all the walkers
{\em up to} step $n$. In this section, we will see that for case I where
$\sigma^2$ is finite, indeed there is a close relation between the
two random variables $R_{n,N}$ and $Y_{n,N}$.  

To uncover this relation, it is actually instructive to consider first the case of 
$N$ independent lattice 
random 
walks defined by Eq. (\ref{markov.2}) where the noise 
$\eta_i(m)$'s are i.i.d. random variables with a distribution 
$f(\eta) = \frac{1}{2} \delta(\eta-1) + \frac{1}{2} \delta(\eta+1)$.
Consider now the time evolution of the two random processes $R_{n,N}$
and $Y_{n,N}$. 
At the next time step $(n+1)$,  
if a new site on the positive axis is visited by any of the walkers 
for the first time, the process $Y_{n,N}$ increases 
by $1$, otherwise its value remains unchanged. Whenever this event happens, i.e.,
a new site on the positive side is visited for the first time, one also
has a record event, i.e., the process $R_{n,N}$ also increases by $1$.
Otherwise $R_{n,N}$ remains unchanged.     
Thus, the
two random processes $Y_{n,N}$ and $R_{n,N}$ are completely locked
with each other at all steps: whenever one of them increases by unity at a given step the 
other does the same simultaneously
and when one of them does not change, the other also remains unchanged. In other
words, for every realization, we have, $Y_{n+1,N}-Y_{n,N}= R_{n+1,N}-R_{n,N}$.
Now, initially all walkers start at the origin indicating $Y_{0,N}=0$ while
$R_{0,N}=1$ since the initial point is counted as a record by convention.
This allows us to write the following identity for all $n$ and $N$ 
\begin{eqnarray}\label{identity}
R_{n,N} = Y_{n,N}+1=\max_{0 \leq m \leq n}\max_{0 \leq i \leq
N} [x_i(m)] + 1 \;.
\end{eqnarray}

We can now take advantage of this identity to compute the probability 
$P(R_{n,N} = M,n)$ as the distribution of $Y_{n,N}$, i.e., the maximum of 
$N$ independent lattice walkers up to $n$ steps can be computed 
using the standard method of images. One obtains, after 
some manipulations left in Appendix \ref{app_lattice}, for $0 \leq M \leq N+1$
\begin{eqnarray}\label{exact_Bernoulli}
P(R_{n,N} = M,n) = \frac{1}{2^{n\,N}} 
\left(\sum_{k=0}^{\lfloor \frac{n+M}{2}\rfloor} \left[ {n \choose k} - {n \choose 
k-M}\right] 
\right)^N - \frac{1}{2^{n\,N}} \left(\sum_{k=0}^{\lfloor \frac{n+M-1}{2} \rfloor}  
\left[{n \choose k} - {n \choose k-M+1}\right] \right)^N \;, 
\end{eqnarray}
where $\lfloor x \rfloor$ is the largest integer not greater than $x$. For instance, for $N=1$ one gets from Eq. (\ref{exact_Bernoulli})
\begin{eqnarray}\label{exact_Bernoulli_N1}
P(R_{n,N} = M,n) = \frac{1}{2^n} {n \choose \lceil \frac{n+M-1}{2}\rceil } \;,
\end{eqnarray} 
where $\lceil x \rceil$ is the smallest integer not less than $x$. 
We have checked that this formula for $N=1$ (\ref{exact_Bernoulli_N1}) 
yields back the result for the first moment $\langle R_{n,1}\rangle$ 
as obtained in Ref. \cite{MZrecord}. From the above 
formula (\ref{exact_Bernoulli}) one can also compute $\langle  R_{n,N} \rangle$, 
for instance with Mathematica, although obtaining a simple closed form formula for it 
for $N>1$ seems rather difficult. In Table \ref{first_values} we have reported the first 
few values of $\langle R_{n,N}\rangle$ for $N=1$ to $N=4$. 
\begin{table}[h]
\begin{center}
\begin{tabular}{|c|c|c|c|c|c|}
  \hline
  \quad & n = 0 & n=1 & n =2 & n = 3 & n = 4 \\
  \hline 
  $\quad$ & $\quad$ & $\quad$ & $\quad$ & $\quad$ & $\quad$  \\
 $N=1$ & 1 & $\dfrac{3}{2}=1.5$ &  $\dfrac{7}{4}=1.75$ &  2 & $\dfrac{35}{16} = 2.187...$ \\
 $\quad$ & $\quad$ & $\quad$ & $\quad$ & $\quad$ & $\quad$  \\
 $N=2$ & $1$ & $\dfrac{7}{4} = 1.75$ & $\dfrac{35}{16}=2.187...$ & $\dfrac{81}{32}=2.531...$ & $\dfrac{723}{256}=2.824...$ \\
$\quad$ & $\quad$ & $\quad$ & $\quad$ & $\quad$ & $\quad$  \\
 $N=3$ & $1$ & $\dfrac{15}{8} = 1.875$ & $\dfrac{157}{64} = 2.453...$ & $\dfrac{731}{256} = 2.855...$ & $\dfrac{13145}{4096} = 3.209...$ \\
 $\quad$ & $\quad$ & $\quad$ & $\quad$ & $\quad$ & $\quad$  \\
 $N=4$ & 1 & $\dfrac{31}{16} = 1.937...$ & $\dfrac{671}{256} = 2.621...$ & $\dfrac{6303}{2048} = 3.077...$ & $\dfrac{227343}{65536} = 3.468...$ \\
 $\quad$ & $\quad$ & $\quad$ & $\quad$ & $\quad$ & $\quad$  \\
  \hline
\end{tabular}
\end{center}
\caption{First values of $\langle R_{n,N}\rangle$ obtained from Eq. (\ref{exact_Bernoulli}).}\label{first_values}
\end{table}

Using the identity (\ref{identity}), one can also obtain the large $n$ behavior of 
$R_{n,N}$. 
Indeed, in this limit, each rescaled ordinary random walk  
$x_i(\tau n)/\sqrt{n}$ converges, when $n \to \infty$, 
to a Brownian motion $B_{D=\frac{1}{2},i}(\tau)$ with a diffusion coefficient $D=1/2$, on the unit time interval, $\tau \in [0,1]$. Therefore from the above identity (\ref{identity}) one gets, in the limit $n \to \infty$
\begin{eqnarray}\label{large_n_discrete}
\frac{R_{n,N}}{\sqrt{n}} \to \tilde M_{N} = \max_{1\leq i \leq N} 
\max_{0 \leq \tau \leq 1} \left[ B_{D=\frac{1}{2},i}(\tau)\right] \;.
\end{eqnarray} 
Now, the distribution of $\max_{0 \leq \tau \leq 1} B_{D=\frac{1}{2},i}(\tau)$, i.e., 
the 
maximum of a single Brownian motion (with diffusion constant $D=1/2$) over a unit 
interval is well known (see e.g., in ~\cite{SMreview})
\begin{eqnarray}\label{discrete_N1}
{\rm Proba.}\,[\tilde M_{1}\leq m] = \sqrt{\frac{2}{{\pi}}} 
\int_0^m\exp\left(-\frac{x^2}{2} 
\right) \; dx = {\rm erf}\left(\frac{m}{\sqrt{2}}\right) \;,
\end{eqnarray} 
Eq. (\ref{large_n_discrete}) demonstrates that in this case, $R_{n,N}/\sqrt{n}$ is distributed like the maximum of a collection 
$N$ i.i.d positive random variables $\{z_1,z_2,\ldots, z_N\}$, each 
drawn from the
distribution: $p(z)= \sqrt{\frac{2}{{\pi}}}\,e^{-z^2/2}$ for $z\ge 0$
and $p(z)=0$ for $z<0$. From Eq. (\ref{discrete_N1}) one obtains also that $\langle R_{1,n}\rangle \approx \sqrt{2n/\pi}$, for $n \gg 1$, as obtained in Ref. \cite{MZrecord}, using a different method. More generally for any $N$ one has
\begin{eqnarray}
\lim_{n \to \infty} {\rm Proba.}\,\left[\frac{R_{n,N}}{\sqrt{n}}\leq m \right] = \left( 
{\rm 
Proba.}\,[\tilde M_{1}\leq m] \right)^N = 
\left[ {\rm erf} \left(\frac{m}{\sqrt{2}}\right) \right]^N \;.
\end{eqnarray}

Having discussed the lattice random walk, let us now return to the 
case where the jump distribution is continuous in space but with a finite $\sigma^2$.
In this case, we do not have an identity between $R_{n,N}$ and $Y_{n,N}$ analogous
to Eq. (\ref{identity}) for lattice random walks. Nevertheless, we conjecture below
and later provide numerical evidence in section VA, that for large $n$, the
scaled random variable $R_{n,N}/\sqrt{n}$ converges, in distribution, to
the scaled variable $Y_{n,N}/\sqrt{n}$ up to a prefactor $\sigma/\sqrt{2}$, i.e.,
\begin{equation}
\lim_{n\to \infty} \frac{R_{n,N}}{\sqrt{n}}\equiv 
\lim_{n\to \infty}\,\frac{\sqrt{2}}{\sigma}\, 
\frac{Y_{n,N}}{\sqrt{n}}
\label{converge1}
\end{equation}
where $\equiv$ indicates that the random variables on both sides of Eq. (\ref{converge1})
have the same probability distribution. On the other hand, using central limit
theorem, it is easy to see that the position of each rescaled walker 
$\frac{\sqrt{2}}{\sigma}\, \frac{x_i(\tau n)}{\sqrt{n}}$ converges in distribution, as 
$n\to \infty$,
to a continuous-time Brownian motion $B_{D=1,i}(\tau)$ with a diffusion coefficient 
$D=1$, over the unit interval $\tau\in [0,1]$. Thus the conjecture in Eq. 
(\ref{converge1}) is equivalent to 
\begin{equation}
\lim_{n\to \infty} \frac{R_{n,N}}{\sqrt{n}}\equiv
M_N= \max_{1\le i\le N} \max_{0 \leq \tau \leq 1} \left[B_{D=1}(\tau)\right] \, .
\label{large_n_continuous}
\end{equation}

The argument leading to this conjecture in Eq. (\ref{large_n_continuous}) can be
framed as follows. Consider first the case for $N=1$.
In this case, the full distribution of $R_{n,1}$ was computed in
Ref. \cite{MZrecord} for all $n$ and in particular, for large $n$, it was
shown that 
\cite{MZrecord} that
 \begin{eqnarray}\label{continuous_N1}
\lim_{n\to \infty} \frac{R_{n,1}}{\sqrt{n}} \equiv M_1 = \max_{0 \leq \tau \leq 1} 
\left[B_{D=1}(\tau)\right] \;,
 \end{eqnarray}
where $B_{D=1}(\tau)$ is a Brownian motion with diffusion coefficient $D=1$ on 
the unit time interval starting from $B_{D=1}(0)$. 
This result (\ref{continuous_N1}), for 
continuous jump distribution, is very similar to the 
one obtained for a lattice random walk in Eq.~(\ref{large_n_discrete}) 
for $N=1$, except that the diffusion coefficient of the Brownian motion 
involved in the discrete case is $D=1/2$ while it is $D=1$, 
independently of $\sigma^2$, for continuous jump distributions. 
In particular, from Eq. (\ref{continuous_N1}) one obtains
\begin{eqnarray}
{\rm Proba.} \, \left(\frac{R_{n,1}}{\sqrt{n}} \leq x\right) \xrightarrow[n\to
\infty]{} {\rm erf}\left(\frac{x}{2}\right) \;.
\end{eqnarray}   

Hence, at least for $N=1$, we know that for large $n$, $R_{n,1}/\sqrt{n}$ for
the continuous jump distribution in Eq. (\ref{continuous_N1}) behaves in a statistically similar way as that
for lattice walks, the only difference is that the effective diffusion constant of
the underlying Brownian motion is $D=1$ in the continuous case (Eq. 
(\ref{large_n_continuous})), while it is $D=1/2$
for the lattice case (Eq. (\ref{large_n_discrete})). Based on this exact relation for 
$N=1$, it is then natural
to presume that this asymptotic equality in law between record numbers for 
continuous jump process and lattice walks
holds even for $N>1$, thus leading to the conjecture in Eq. (\ref{large_n_continuous}).
As a first check of the validity of this conjecture, we note that the result
for the first moment of the record number in Eqs. (\ref{avgrec.1}) and (\ref{maximum.2}) 
is fully consistent with the conjecture in Eq. (\ref{large_n_continuous}). 

As announced in the introduction (\ref{gumbel.2}), the conjecture in 
Eq. (\ref{large_n_continuous}) is equivalent to say that $R_{n,N}/\sqrt{n}$ converges, 
in distribution, to the maximum of $N$
independent random variables $M_N= {\rm max}(y_1,y_2,\ldots, y_N)$, 
where $y_i\ge 0$'s are i.i.d non-negative random variables each drawn from  
distribution $p(y)= \frac{1}{\sqrt{\pi}}\, e^{-y^2/4}$ for $y\ge 0$ and $p(y)=0$ for 
$y<0$. In particular, the cdf of $R_{n,N}/\sqrt{n}$ is given by 
\begin{eqnarray}\label{conjecture}
{\rm Prob.} \left( \frac{R_{n,N}}{\sqrt{n}} \leq x\right) \xrightarrow[n\to
\infty]{} \left[ {\rm erf}\left(\frac{x}{2}\right) \right]^N \;.
\end{eqnarray}
In section VB we will demonstrate that this conjecture is well supported by our numerical 
simulations. From Eq. 
(\ref{large_n_continuous}), one can then use standard results for the 
extreme statistics of independent random variables \cite{gumbel} 
to obtain that for large $N$ and large $n$, the scaled random variable 
$R_{n,N}/\sqrt{n}$ (properly shifted and scaled) converges, in distribution, 
to the Gumbel distribution as announced in Eq. (\ref{gumbel.1}). 

Although we have not found a rigorous proof of the above result 
(\ref{large_n_continuous}), the fact that both formulae for random walk with discrete 
(\ref{large_n_discrete}) and continuous (\ref{large_n_continuous}) jump distribution 
differ essentially by a factor of $\sqrt{2}$ is reminiscent of a similar difference, by 
the same factor of $\sqrt{2}$, for the survival probability $q_n(0)$ corresponding to 
both random walks (starting from the origin). This quantity $q_n(0)$ plays indeed a 
crucial role in the computation of the record statistics of a random walk 
\cite{MZrecord}. For the lattice random walk, one has indeed $q_n(0) \sim \sqrt{2/\pi} 
\, n^{-1/2}$ while for the continuous random walk one has, from Sparre-Andersen's 
theorem, $q_n(0) \sim \sqrt{1/\pi}\, n^{-1/2}$, independently of $\sigma^2$. This fact 
certainly deserves further investigations.

In the case of divergent $\sigma^2$, the two random variables
$R_{n,N}$ and $Y_{n,N}$ do not seem to be related in any simple way.
This is already evident from 
the result for the mean record number $\langle 
R_{n,N}\rangle$ 
in Eq. (\ref{avgrec.case2}) for $0\le \mu <2$. 
As $n\to \infty$, $\langle R_{n,N}\rangle \approx 4/\sqrt{\pi}\, \sqrt{n}$ 
where the prefactor does not depend on $N$ for large $N$.
In contrast, using standard extreme value statistics~\cite{gumbel}, 
it is easy to show that the mean value $\langle Y_{n,N}\rangle \sim (N\, n)^{1/\mu}$
for large $n$ and $N$ with $0<\mu<2$. This rather different asymptotic behavior of
the mean thus already rules out any relationship between $R_{n,N}$ and $Y_{n,N}$
for case II. So, for this case, our result for the distribution of $R_{n,N}$
is only restricted to numerical simulations that are presented in the next section.  

\section{Numerical simulations}

In this section we present the results of our numerical simulations of $N$ independent 
random walks both with a finite $\sigma^2$ (case I) and with a divergent $\sigma^2$ (case 
II) and compare them with our analytical results. In the first subsection 
\ref{subsection_distributions} we study the statistics of the record numbers (both its 
mean value and its full distribution). Since, at least in case I, the mean record number 
is strongly related to the expected maximum of the process we will then analyze the 
evolution of the largest of the $N$ random walkers. This will be done in section 
\ref{subsection_max}. We find that the statistics of the maximum significantly differs 
between the cases I and II. Finally, in section \ref{subsection_correlations} we will 
consider the correlations between individual record events in the two different cases and 
show that, at least asymptotically, these correlations are not different from the case of 
only one single random walker.

\subsection{Statistics of the record numbers}
\label{subsection_distributions}

\begin{figure}[h]
\includegraphics[angle=-90,width=0.6\textwidth]{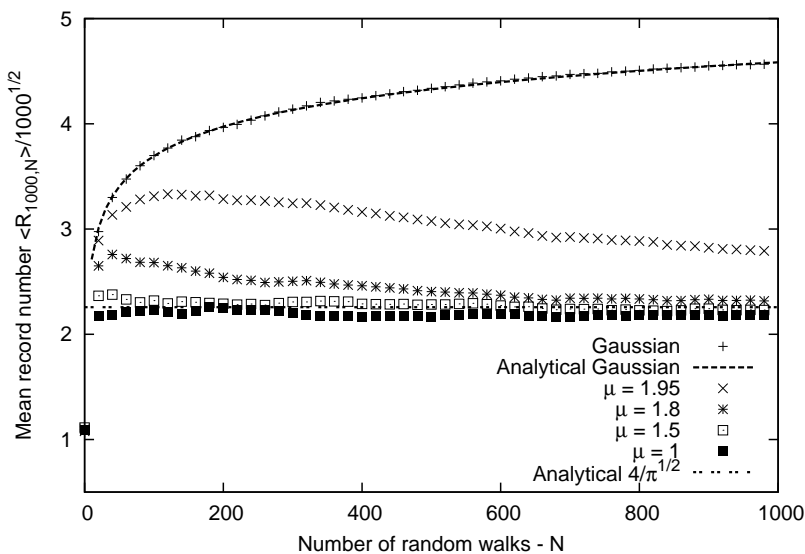}
\caption{Rescaled mean record number $\langle R_{n,N}\rangle / \sqrt{n}$ for a 
fixed $n=1000$ plotted against the number of random walkers $N$. 
We performed simulations with jump distributions of the type 
$f\left(\eta\right) \sim |\eta|^{-\mu-1}$ and different $\mu = 1, 1.5, 1.8$ and $1.95$ 
and for the Gaussian jump distribution with zero mean and $\sigma^2=1$. The Gaussian case 
is compared to our analytical 
finding for the finite $\sigma^2$ case (Eq. (\ref{avgrec.case1})), which 
is given by the dashed line. The thick black line gives the analytical result for the infinite $\sigma^2$ regime (Eq. (\ref{avgrec.case2})). With increasing $N$ all $\langle R_{n,N}\rangle / \sqrt{n}$ with $\mu<2$ approach this line of value $4/\sqrt{\pi}$.}
\label{rec_no.fig} 
\end{figure}

{\bf Case I ($\sigma^2$ finite)}. 
In Fig.  (\ref{rec_no.fig}), we show our 
numerical results for $\langle R_{n,N} \rangle$ for $\sigma^2$ finite, 
which were obtained by a direct simulation of the jump process in Eq. (\ref{markov.1}) 
with $n = 10^4$ 
steps, with a Gaussian distribution of the jump variables $\eta_i(m)$'s
(mean $0$ and $\sigma^2=1$). These results have been obtained by averaging over $10^3$ 
different realizations of 
the random walks. These data, on Fig.  (\ref{rec_no.fig}) are indexed by the label 'Gaussian'. Our numerical data show a very nice agreement with our analytical result obtained in Eq. (\ref{avgrec.1}) yielding $\langle R_{n,N}\rangle / \sqrt{n} = \alpha_N$. The large $N$ behavior of $\alpha_N$ can be easily obtained by a saddle point analysis, yielding:
\begin{eqnarray}\label{sub_leading}
\frac{\langle R_{n,N}\rangle}{\sqrt{n}} = 2 \sqrt{\log N} - \frac{\log(\log N)}{2 \sqrt{\log N}} + {\cal O}[(\log N)^{-1/2}] \;.
\end{eqnarray}    
It turns out that for $N\sim1000$, the sub-leading corrections (\ref{sub_leading}) are still sizeable.  

\begin{figure}[h]
\includegraphics[angle=-90,width=0.6\textwidth]{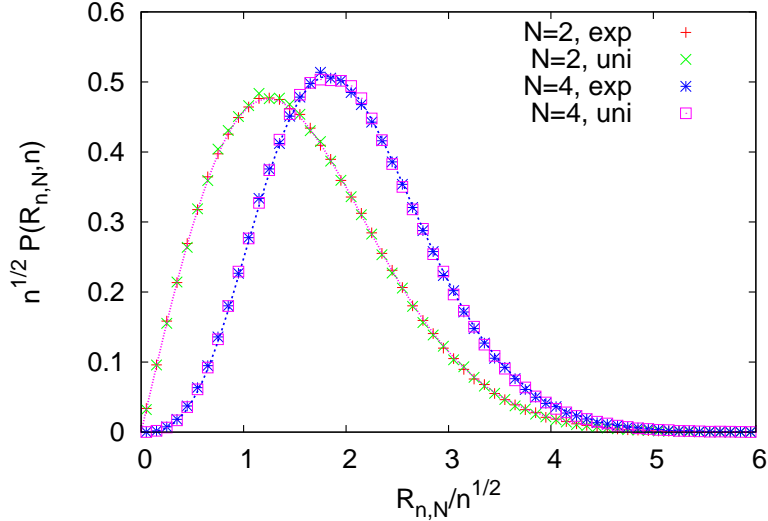}
\caption{Scaled probability distribution function $\sqrt{n} P(R_{n,N},n)$ as a function of $R_{n,N}/\sqrt{n}$ for $N=2$ and $N=4$ independent of random walks of length $n = 10^4$. For each value of $N$, we show the result for the case where the jumps are distributed exponentially ('exp') and uniformly between $-1/2$ and $1/2$ ('uni'). The dotted lines correspond to the result in Eq. (\ref{conjecture}) which we conjectured to be the exact one. There are no fitting parameters.}\label{check_conjecture}
\end{figure}

We have also computed numerically the distribution of the (scaled) record numbers $R_{n,N}/\sqrt{n}$ and compared it to our conjecture in Eq. (\ref{large_n_continuous}). The results of this comparison, for different values of $N=2, 4$ are shown in Fig. \ref{check_conjecture} ,where the data were obtained by averaging over $5 \times 10^4$ realizations of independent random walks of $n=10^4$ steps. The data, for two different continuous jump distributions (exponential and uniform) show a very nice agreement with our analytical prediction in Eq. (\ref{conjecture}), which we argue to be an exact result. As mentioned above, one expects that this distribution will converge, for $N \to \infty$ to a Gumbel distribution (\ref{gumbel.1}), albeit with strong finite $N$ effects.

{\bf Case II ($\sigma^2$ divergent)}. In Fig. (\ref{rec_no.fig}), we show our numerical results for $\langle R_{n,N} \rangle$ for $\sigma^2$ divergent, obtained by a direct simulation of the random walk (\ref{markov.1}) where the distribution of $\eta_i(m)$'s has a power law tail $f(\eta) \sim |\eta|^{-1 - \mu}$ with $\mu < 2$, and different values of $\mu$. The data presented there were also obtained by averaging over $10^3$ different realizations of random walks, with $10^4$ steps. These data show that, in this case, $\langle R_{n,N}\rangle/\sqrt{n}$ approaches a constant value for fixed (and large) $n$ and $N\rightarrow\infty$, which is fully consistent with the value of $4/\sqrt{\pi}$ obtained analytically in Eq. (\ref{avgrec.case2}). The simulations also show how the speed of this convergence is modified when $\mu<2$ is varied. While for small $\mu\ll2$, $\langle R_{n,N}\rangle / \sqrt{n}$ approaches the universal limit value of $4/\sqrt{\pi}$ very quickly, we find a slower convergence for $2-\mu\ll 1$.

\begin{figure}[h]
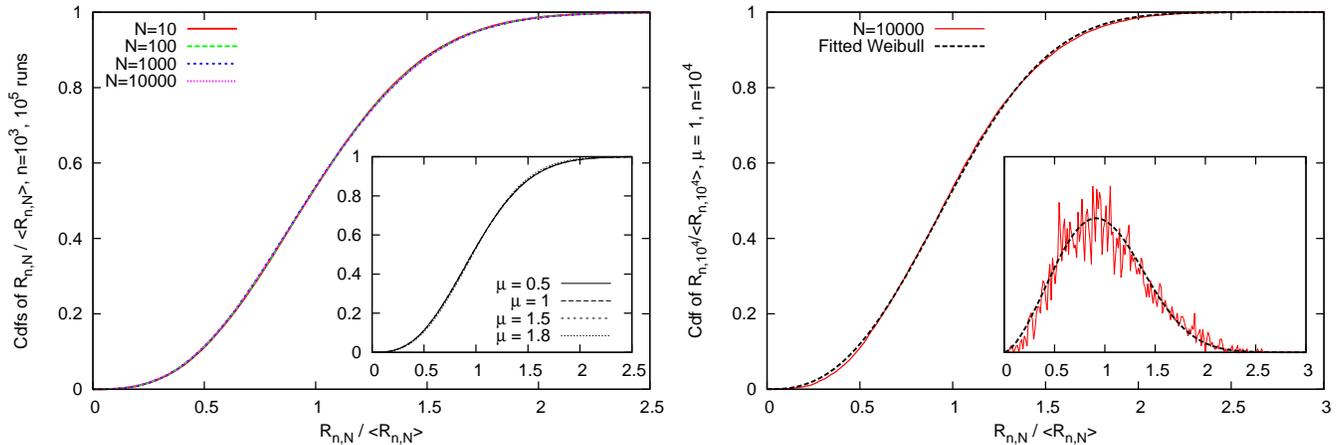

\includegraphics[angle=-90,width=0.5\textwidth]{distribution1.eps}\includegraphics[angle=-90,width=0.5\textwidth]{distribution2.eps}
\caption{\textbf{Left:} Cumulative distribution function (cdf) of the scaled variable $R_{n,N}/\sqrt{n}$ for the L\'evy index $\mu=1$, which approaches the universal distribution $F_2\left(x\right)$. The figure gives results for a fixed $n=10^3$ and different $N$, for each $N$ we performed $10^5$ simulations. The inset gives simulations of the cdf for fixed $N=10^3$ while L\'evy index is varied. \textbf{Right:} The cdf for $\mu=1$ and $N=10^4$. We have fitted the data with a Weibull distribution as in Eq. (\ref{Weibull}), where the fitting parameters were $\lambda \approx 0.8944 \pm 0.0003$ and $k = 2.558 \pm 0.003$. The inset gives the corresponding curves for the pdf's.}
\label{distributions.fig} 
\end{figure}

The numerical computation of the distribution of the (scaled) record number $R_{n,N}/\sqrt{n}$ is of special interest because an analytical study of it, beyond the first moment, is still lacking. The two plots on Fig. (\ref{distributions.fig}) show our numerical results for this distribution, which were obtained by averaging over $10^4$ independent random walks of length $n=10^4$. 
The left panel in Fig. (\ref{distributions.fig}) shows the rescaled distribution of 
$R_{n,N}/\sqrt{n}$ at a fixed time step $n=10^3$ and $\mu=1$. Apparently all curves for 
$N=10,10^2,10^3$ and $10^4$ collapse on one line. In the inset of the left figure we kept 
$n=10^3$ and $N=10^3$ fixed and varied $\mu$: one can see that all the cdf's collapse. 
These results suggest that \begin{equation}
{\rm Prob.}\left[ \frac{R_{n,N}}{\sqrt{n}}\le x\right] \xrightarrow[N\to
\infty]{n\to
\infty} F_2(x) \;,
\label{distri2.1}
\end{equation}
where the limiting distribution function $F_2\left(x\right)$ is independent of $\mu<2$. We tried to guess the analytic form of $F_2\left(x\right)$ by comparing with several known continuous distributions that are defined for positive real numbers. We are certain that $F_2\left(x\right)$ is not a Gaussian distribution. By far the best results were obtained by fitting with a Weibull distribution 
\begin{equation}
\label{Weibull}
F_2\left(x\right) = 1-\textrm{exp}\left(-\left(\lambda x\right)^k\right) \;,
\end{equation}
with two free real parameters $\lambda>0, k>0$. Fitting with the least-squares method gives values of $\lambda \approx 0.8944 \pm 0.0003$ and $k = 2.558 \pm 0.003$. The right panel in Fig. (\ref{distributions.fig}) compares this fit with the distribution obtained from a simulation of $N=10^4$ random walks of length $n=10^4$. While the agreement, both for the cdf and the probability density function (pdf) is quite good, there are still some deviations between the two, particularly for small values close to zero. We are not sure, if this difference is a finite $N$ effect or if the real limit distribution of $R_{n,N}/\sqrt{n}$ for $N\rightarrow\infty$ effectively differs from a Weibull distribution.


\subsection{Temporal evolution of two stochastic processes: the record number $R_{n,N}$
and the global maximum $Y_{n,N}$ up to step $n$} 
\label{subsection_max}

\begin{figure}
\includegraphics[angle=-90,width=0.5\textwidth]{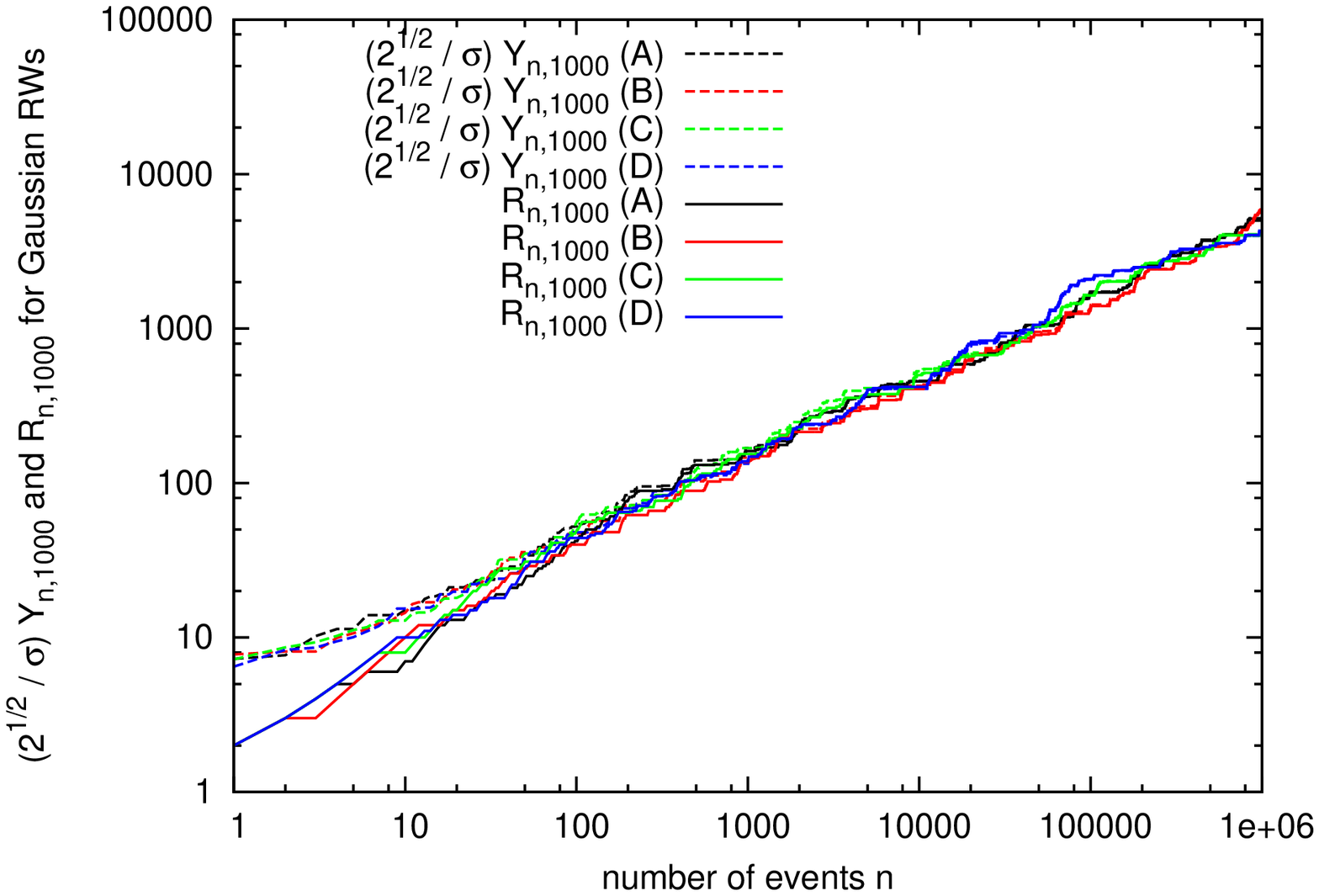}\includegraphics[angle=-90,width=0.5\textwidth]{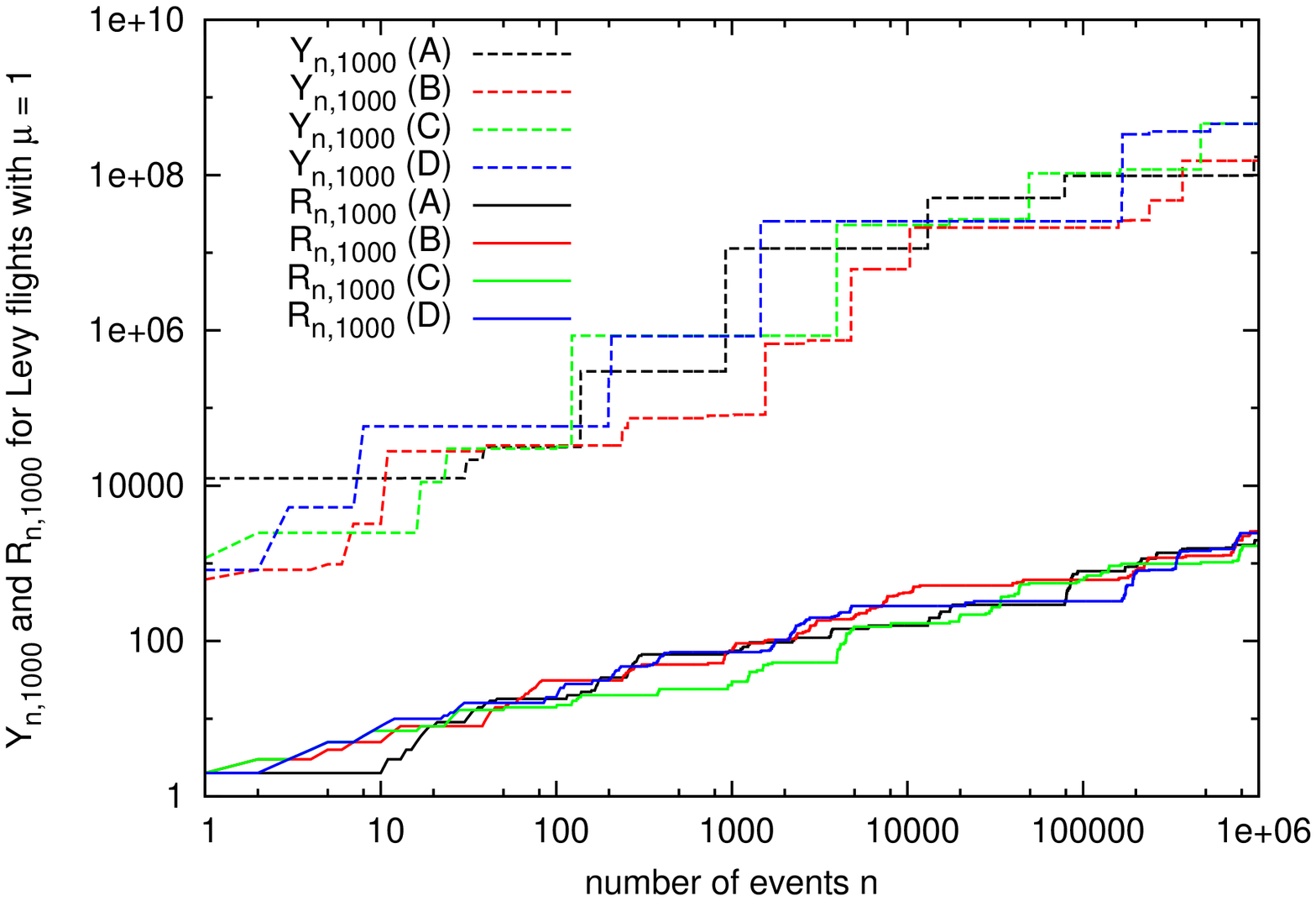}
\caption{\textbf{Left:} Time evolution of the record number $R_{n,N}$ and 
the rescaled maximum value $(\sqrt{2}/\sigma)\, Y_{n,N}$ reached up to 
the $n$-th step for 
four different realizations of $N=1000$ random walks (labeled A,B,C,D) with 
Gaussian jump distribution (zero mean and $\sigma=1$) (case I). Here, the results look 
rather 
deterministic and for $n\rightarrow\infty$, we find $(\sqrt{2}/\sigma)\, Y_{n,N} \approx 
R_{n,N}$ for every realization. \textbf{Right:} $R_{n,N}$ and $Y_{n,N}$ for four 
different realizations of 
$N=1000$ L\'evy flights (labeled A,B,C,D) with $\mu=1$. The behavior of 
$Y_{n,N}$ for the L\'evy flight is completely different from $R_{n,N}$:
while $R_{n,N}$ is self-averaging, $Y_{n,N}$ fluctuates widely from
one sample to another and is not self-averaging.} 
\label{max.fig} 
\end{figure}

In the case of jump distributions with a finite second moment $\sigma^2$ (case I), we 
have shown that the mean $\langle Y_{n,N}\rangle$ of the maximum of all walkers up to 
step $n$ and the mean record number $\langle R_{n,N}\rangle$ are proportional to each 
other in the limit $n\rightarrow\infty$, both grow with $\sqrt{n \ln N}$. In contrast, 
for L\'evy walks with index $\mu<2$ (case II) the relation between these two observables 
does not hold any more and the mean record number grows much slower than the maximum (see 
the discussion at the end of section IV). To illustrate 
the similarities and differences in the growth rate of $R_{n,N}$ and $Y_{n,N}$
in the two cases (I and II),
we compare their respective time 
evolution for $4$ different samples in Fig. (\ref{max.fig}). 
On the left panel, we consider the Gaussian jump distribution with
zero mean and $\sigma^2=1$ and we see that the process $R_{n,N}$ and
$(\sqrt{2}/\sigma)\,Y_{n,N}$ become identical very quickly. Moreover,
the two processes evolve almost in a deterministic fashion with growing $n$
and hardly fluctuate from one sample to another.
In contrast, on the right panel where we plot the two processes for $\mu=1$,
their behavior change drastically. First of all, the two processes $R_{n,N}$
and $Y_{n,N}$ do not seem to have relation to each other. While $R_{n,N}$
again evolves almost deterministically and in a self-averaging way,  
the trajectories of the process $Y_{n,N}$ differ strongly from one sample 
to another and $Y_{n,N}$ is clearly non self-averaging. In particular, the process 
$Y_{n,N}$ can, like in a single L\'evy flight, perform very large jumps exceeding its 
previous value by several orders of magnitude.

\subsection{Correlations between record events}
\label{subsection_correlations}
\begin{figure}
\includegraphics[angle=-90,width=0.6\textwidth]{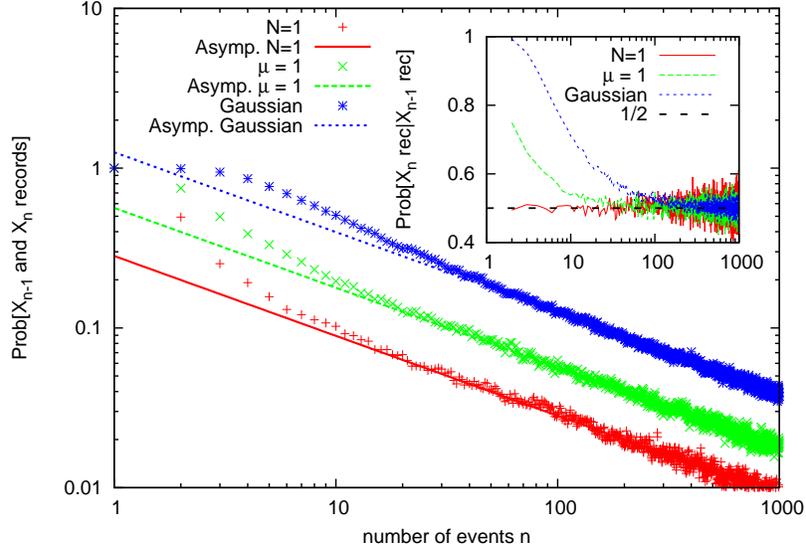}
\caption{Probability for two successive records at times $n-1$ and $n$ for a single random walker as well as $N$ random walks with jump distributions of tail-exponents $\mu=2$ (case 1) and $\mu=1$ (case 2). In all three cases this probability asymptotically approaches a value of $1/2$ times the probability for a record in the $n$th step. Therefore, for large $n$, the correlations between the record events in the $N\gg1$ regime are the same as in the $N=1$ case. This is also shown by the inset, where we plotted the probability for a record in the $n$th step conditioned on a record in the previous one, which approaches $1/2$ in all three cases.}
\label{correlations.fig} 
\end{figure}

An important feature of the record statistics of a single random walk ($N=1$) is its renewal property, which leads to the fact, that each time after a record event, the record statistics is, in some sense, \textit{reseted}. After a record event the process evolves as a new process with the record value as its new origin. Therefore it is very simple to give the pairwise correlations between record events. In fact, from the above argument, we have
\begin{equation}
\textrm{Prob}\left[\textrm{rec. at}\;n-k\;\textrm{and}\;n\right] = \textrm{Prob}\left[\textrm{rec. at}\;n-k\right] \times \textrm{Prob}\left[\textrm{rec. at}\;k\right] = r_{n-k,1}r_{k,1} \;.
\end{equation}
Using the results from \cite{MZrecord} this gives the following (exact) result for $\textrm{Prob}\left[\textrm{rec. at}\;n-k\;\textrm{and}\;n\right]$: 
\begin{equation}
\textrm{Prob}\left[\textrm{rec. at}\;n-k\;\textrm{and}\;n\right] = \left(2\left(n-k\right) \atop n-k\right) \left(2k\atop k\right) 2^{-2n}  \;.
\end{equation}
In the special case of $k=1$ we find $\textrm{Prob}\left[\textrm{rec. at}\;n-1\;\textrm{and}\;n\right] = \frac{1}{2} r_{n-1}$. With this we find that the conditional probability of a second record directly following a record that just occurred is always given by
\begin{equation}
 \textrm{Prob}\left[\textrm{rec. at}\;n|\textrm{rec. at}\;n-1\right] = \frac{1}{2} \;.
\end{equation}
In our efforts to understand and compute the distribution of the record number $R_{n,N}$ for L\'evy flights (case II), we considered the correlations between successive record events also for $N\gg1$ random walks. If the correlations between successive record events in the large $N$ limit would vanish, one could assume that the asymptotic distribution of $R_{n,N}$ approaches a Gaussian. However, we found that this is not the case. Fig. (\ref{correlations.fig}) gives the behavior of $\textrm{Prob}\left[\textrm{rec. at}\;n-k\;\textrm{and}\;n\right]$ for the $N=1$ case, as well as for $N=10^3$ for random walks of the two cases I and II with L\'evy indices $\mu=2$ and $\mu=1$. In all three cases $\textrm{Prob}\left[\textrm{rec. at}\;n-1\;\textrm{and}\;n\right]$ approaches $\frac{1}{2} \textrm{Prob}\left[\textrm{rec. at}\;n-1\right] = \frac{1}{2}r_{n,N}$ for large $n$, proving that for $n\rightarrow\infty$ the probability for a second record after an occurred one is just $1/2$. The inset of Fig. (\ref{correlations.fig}) also shows this behavior. Here, while for $N=1$ the conditional probability for a second record is always $1/2$, this value is only approached for larger $n$ in the case of $N\gg1$. For small $n$ the conditional probability is larger.

\section{Comparison to stock prices} 
\label{section_stocks}
The oldest application of the random walk model, which was already proposed by Le Bachelier \cite{LB} in 1900, is the one to stock data \cite{MantegnaStanley,Voit}. In his model the stock prices perform a so-called geometric random walk and trends in the stocks are modeled by a linear drift in the logarithms of the stock prices. In \cite{WBK} the record statistics of stocks in the Standard and Poors 500 \cite{SP500} (S\&P 500) index were compared to the records in a random walk with a drift. The authors could show that on average, on a time interval of $n=100$ trading days, the statistics of upper records in individually detrended stocks are in good agreement with the same statistics of a random walk with a symmetric jump distribution. The lower records however showed a significant deviation from this model. 

Here, we want to extend this analysis to the record statistics of $N$ stocks. The question is, to what degree, the record statistics of $N$ normalized and randomly chosen stocks from the S\&P 500 can be compared to the record statistics of $N$ independent random walks. As in \cite{WBK}, the observational data we used consisted of $366$ stocks that remained in the S\&P 500 index for the entire time-span from January 1990 to March 2009. Overall, we had data for $5000$ consecutive trading days for each stock at our disposal. In \cite{WBK} we found that it is useful to analyze this data over smaller intervals, on which one can then detrend the measurements. We decided to split up the $5000$ trading days into $20$ consecutive intervals of each $250$ trading days, which is roughly one calendar year. In each of these intervals we considered the logarithms $X_i = \ln S_i / S_0$ of the stock prices $S_i$, where $S_0$ is the first trading day. The random walk model then suggest that these logarithms $X_i$ perform a biased random walk that starts at the origin ($X_0=0$). Since our analytical theory presented in this paper only works for symmetric random walks we had to detrend the stocks. We subtracted an index-averaged linear trend obtained by linear regression from the $X_i$'s in order to obtain symmetric random walkers. Finally, in order to make the stocks comparable to our model of $N$ random walkers of the same jump distribution, we had to normalize the $X_i$'s by dividing through the standard deviations of the respective individual jump distributions. After this detrending and normalization we can assume that the jump distributions in the individual time series have at least the same mean and the same variance, which should then be given by 0 and 1.

\begin{figure}
\includegraphics[angle=-90,width=0.5\textwidth]{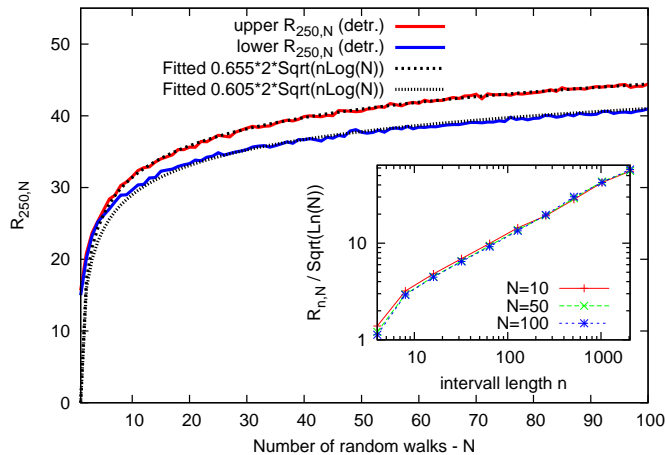}
\caption{The averaged upper and lower record number after $n=250$ trading days in the S\&P 500 stock data. The $5000$ trading days in \cite{SP500} were subdivided in $20$ intervals of $250$ days and then linearly detrended in these intervals using the average linear trend of the index. Then we chose $N$ stocks randomly out of the total number of $366$ stocks and analyzed the evolution of the record number in this set. This random picking was repeated $10^4$ times and the results were averaged to obtain the figure. The dashed lines give our analytical prediction for $N$ Gaussian random walks multiplied by fitted prefactors. The inset gives the behavior of the $\langle R_{n,N}\rangle/\sqrt{\ln N}$ for different $N$ plotted against the interval length $n$, confirming the proportionality $\langle R_{n,N}\rangle \propto \sqrt{\ln N}$. }
\label{stocks.fig} 
\end{figure}

For a fixed $N$, we randomly selected subsets of size $N$ from the $366$ detrended and normalized stocks for each of the $20$ intervals of length $n=250$ and computed the evolution of the record number $R_{n,N}$ in these subsets. To get reliable statistics we average $R_{n,N}$ over $10^4$ different subsets with a fixed $N$ and also averaged over the $20$ consecutive intervals. The resulting $\langle R_{n,N}\rangle$'s for the upper and the lower record number and $N$ between $1$ and $100$ are given in Fig. (\ref{stocks.fig}). We find that both the curve for the upper and the curve for the lower mean record number are not in agreement with our theoretical prediction for the case of $N$ Gaussian random walks given by $\langle R_{n,N}\rangle = 2\sqrt{n\ln N}$. However, Fig. (\ref{stocks.fig}) shows, that the $\langle R_{n,N}\rangle$ for the stocks increase with $N$. We also considered subsets of size $N>100$ and found that for $N$ closer to the maximal value of $366$, $\langle R_{n,N}\rangle$ gets almost constant in $N$.
%
%
While the increase of $\langle R_{n,N}\rangle$ for smaller $N$ indicates that the statistics behave like $N$ independent Gaussian random walks, the fact that it saturates for large $N$ could indicate that they behave like a L\'evy flight with L\'evy index $\mu<2$. We know however, that the tail exponent of the daily returns $\ln S_{i} / S_{i-1}$ in the stock data is much larger than $\mu=2$ and that they definitely do not perform a L\'evy flight \cite{GGPS2003,PGAMS2008,GWunpup}. Much more likely is that the correlations between the individual stocks play an important role. In addition, we observed that at least for $N<100$, $\langle R_{n,N}\rangle$ grows proportional to $2\sqrt{n\ln N}$. One way to interpret this finding is the following: When we assume that in $N$ stocks only a smaller number of $N^{\gamma}$ (with $\gamma\in\mathbb{R}^+$ and $\gamma<1$) is effectively independent and that only these $N^{\gamma}$ stocks contribute to the record statistics, the mean record number should be given by
\begin{equation} 
\langle {R_n,N}\rangle = \langle {R_n,N^{\gamma}}\rangle^{\left(\textrm{Gaussian}\right)} = 2 \sqrt{\gamma\; n\ln N} \;,
\end{equation}
and saturate if the value of $N_{max}^{\gamma}$ is achieved, where $N_{max}$ is the total number of stocks. In Fig. (\ref{stocks.fig}) we fitted curves of the form $\sqrt{\gamma_{\pm}}\; 2\sqrt{n\ln N}$ with $\sqrt{\gamma_{+}} \approx 0.655$ and $\sqrt{\gamma_{-}} \approx 0.605$ to the development of the upper and lower $\langle R_{n,N}\rangle$. The good agreement with the fitted curves and the data confirms our assumption. Apparently, the record statistics of $N$ detrended and normalized stocks is the same as the one of $N^{\gamma}$ independent Gaussian random walks. This finding is also confirmed by the inset in Fig. (\ref{stocks.fig}). There we plotted $\langle R_{n,N}\rangle / \sqrt{\ln N}$ for different interval length $n$ and some different subset sizes $N$. The fact that all the lines collapse tells us that $\langle R_{n,N}\rangle / \sqrt{\ln N}$ is independent of $N$ and therefore 
\begin{equation}
 \langle R_{n,N}\rangle \propto \sqrt{n \ln N}.
\end{equation}

\section{Conclusion}

In conclusion we have presented a thorough analysis of the record statistics of $N$ 
independent random walkers with continuous and symmetric jump distributions. For $N > 1$, 
we have found 
two distinct cases: the case where the variance of the jump distribution $\sigma^2$ is 
finite and the case where $\sigma^2$ does not exist (case 
II) as in the case of L\'evy random walkers with index $0 < \mu < 2$. In 
the first case we have found that the mean record number behaves like $\langle R_{n,N} 
\rangle \approx 2 \sqrt{\log N} \sqrt{n}$ for $n, N \gg 1$ while in case II, $\langle 
R_{n,N} \rangle \approx \sqrt{4/\pi} \sqrt{n}$ for $n, N \gg 1$.
 
We have then argued that, in the first case, the full distribution of the scaled number 
of records $R_{n,N}/\sqrt{n}$ is given by the distribution of the maximum of $N$ 
independent Brownian motions with diffusion coefficient $D=1$. This statement was 
suggested by an exact result for lattice random walks and it was corroborated (i) by our 
exact calculation of the first moment $\langle R_{n,N}\rangle/\sqrt{n}$ valid for any 
value of $N$ and (ii) by our numerical simulations. Of course it would be very 
interesting to obtain a proof of this result. From this connection with extreme value 
statistics, one thus expects that the distribution of $R_{n,N}/\sqrt{n}$ converges, for 
$N \to \infty$, to a Gumbel form (\ref{gumbel.1}). This connection between record 
statistics and extreme value statistics could also be useful to compute the record 
statistics in other models discussed in the introduction, for instance, in the Linear 
Drift Model~\cite{FWK1, WK, WBK}. In the case of L\'evy random walkers, 
we have shown numerically that the full distribution of $\langle R_{n,N}\rangle/\sqrt{n}$ 
converges, when $N \to \infty$, to a limiting distribution $F_2(x)$ which is independent 
of $\mu$. The exact computation of this universal distribution remains a challenging 
problem. Other interesting questions concern the extension of these results to include a 
linear drift \cite{PLDW2009} or to the case of constrained L\'evy walks, like L\'evy 
bridges which were recently studied in the context of real space condensation phenomena 
\cite{SM2010}. Finally the applications of our results to the record statistics of stock 
prices from the Standard \& Poors 500 index suggest that, among a set of $N$ stocks, 
only a smaller number, which scales like $N^\gamma$, with $0 < \gamma < 1$, are 
effectively independent. The record statistics of these $N^\gamma$ stocks is then very 
similar to the statistics of $N^\gamma$ independent random walkers. This idea might be 
useful for future investigations of the fluctuations of such ensemble of stock prices.

\vskip 0.3cm \noindent\textbf{Acknowledgments:} We acknowledge support by ANR grant 
2011-BS04-013-01 WALKMAT and the Department of Business Administration and Finance at the 
University of Cologne for providing access to the stock data from the S\&P 500. GW is 
grateful for the kind hospitality of the Laboratoire de Physique Th\'eorique et Mod\`eles 
Statistiques during the completion of this work and for the financial support provided by 
DFG within the Bonn Cologne Graduate School of Physics and Astronomy.

\appendix

\section{Scaling behavior of $p_m(x)$ and $q_m(x)$ for finite $\sigma^2$}

We start from Eqs. (\ref{pm.2}) and (\ref{qm.2}). When $\sigma^2$ is finite,
by central limit theorem, the typical position of a walker after $m$ steps scales as
$m^{1/2}$ for large $m$. Hence the natural scaling variable is $z=x/m^{1/2}$. 
Consider first Eq. (\ref{pm.2}) satisfied by $p_m(x)$. To extract the 
leading scaling function in the scaling 
limit $x\to \infty$, $m\to \infty$ but keeping $z=x/m^{1/2}$ fixed,
we need to investigate $\phi(s,\lambda)$, given explicitly in Eq. 
(\ref{ivanov.2}), in the limit when $\lambda\to 0$, $s\to 1$ but keeping
the ratio $\lambda/\sqrt{1-s}$ fixed. To extract the behavior of 
$\phi(s,\lambda)$ in this scaling limit, it is advantageous to work with an 
alternative expression of $\phi(s,\lambda)$ derived in Ref. \cite{MCZ2006}
for finite $\sigma^2$ 
\begin{equation}
\phi(s,\lambda)= \frac{1}{[\sqrt{1-s}+ \sigma \lambda 
\sqrt{s/2}]}\,\exp\left[-\frac{\lambda}{\pi}\, \int_0^{\infty} 
\frac{dk}{\lambda^2+k^2}\, \ln\left[\frac{1-s {\hat f}(k)}{1-s+ 
s\sigma^2k^2/2}\right]\right].
\label{phisl1}
\end{equation}
This expression is more suitable for extracting the scaling limit. In the limit 
$\lambda\to 0$ and $s\to 1$, the expression inside the exponential in Eq. 
(\ref{phisl1}) tends to $0$ and hence, to leading order, we have
\begin{equation}
\phi(s,\lambda)\approx \frac{1}{[\sqrt{1-s}+ \sigma \lambda
\sqrt{s/2}]}\, .
\label{phisl2}
\end{equation}
Inverting the Laplace transform with respect to $\lambda$ (it has a simple pole
at $\lambda=-\sqrt{2(1-s)}/\sigma$) one gets from Eq. (\ref{pm.2})
\begin{equation} 
\sum_{m=0}^{\infty} s^m p_m(x) \approx \frac{\sqrt{2}}{\sigma}\, 
e^{-\sqrt{2(1-s)}\, x/\sigma}\,.
\label{genpm.1}
\end{equation}
Setting $s=1-p$ with $p\to 0$ in the scaling limit, the sum on the lhs of Eq. 
(\ref{genpm.1}) can be approximated, to leading order, by a continuous integral:
$\sum_{m=0}^{\infty} s^m p_m(x) \approx \int_0^{\infty} p_m(x)\, e^{-pm} dm$
and we have
\begin{equation}
\int_0^{\infty} p_m(x)\, e^{-p\,m}\, dm\approx \frac{\sqrt{2}}{\sigma}\,
e^{-\sqrt{2p}\, x/\sigma}\,.
\label{genpm.2}
\end{equation}
Next we need to invert the Laplace transform with respect to $p$. We use
the explicit inversion formula, ${\cal L}_{p\to m}^{-1}[e^{-b \sqrt{p}}]=
\frac{b}{2\,\sqrt{\pi}\,m^{3/2}}\, \exp[-b^2/4m]$.
Applying this to Eq. (\ref{genpm.2}) gives, to leading order,
in the scaling limit
\begin{equation}
p_m(x) \approx \frac{x}{\sigma^2 \sqrt{\pi}\, 
m^{3/2}}\,\exp\left[-\frac{x^2}{2\sigma^2 m}\right] \:,
\label{pms.1}
\end{equation}
which can be reorganized in the scaling form 
\begin{equation}
p_m(x)\to  \frac{1}{\sqrt{2\sigma^2}\, m}\,
g_1\left(\frac{x}{\sqrt{2\,\sigma^2\,
n}}\right)\,,
\quad
{\rm where}\quad g_1(z)= \frac{2}{\sqrt{\pi}}\,z\, e^{-z^2}\, . 
\label{pms.2}
\end{equation}

Next we consider $q_m(x)$ given in Eq. (\ref{qm.2}). Following exactly the same 
procedure as in the case of $p_m(x)$ we find, in the scaling limit,
\begin{equation}
\int_0^\infty q_m(x)\, e^{-p m}\, dm \approx \frac{1}{p}\left[1- 
e^{-\sqrt{2p}\,x/\sigma}\right].
\label{qms.1}
\end{equation}
Inverting the Laplace transform with respect to $p$ upon using the
explicit inversion formula, ${\cal L}_{p \to m}^{-1}[e^{-b \sqrt{p}}/p]={\rm 
erfc}(b/\sqrt{4m})$, we get, to leading order in the scaling limit
\begin{equation}
q_m(x) \approx 1- {\rm erfc}\left(\frac{x}{\sqrt{2\sigma^2\, m}}\right)={\rm 
erf}\left(\frac{x}{\sqrt{2\sigma^2\,m}}\right) \;,
\label{qms.2}
\end{equation}
which proves the result in Eq. (\ref{qm1scaling}). 

\section{Scaling behavior of $p_m(x)$ and $q_m(x)$ for divergent $\sigma^2$}

We consider jump distribution $f(\eta)$ such that its Fourier transform behaves, 
for 
small $k$, as ${\hat f}(k) \approx 1- |ak|^{\mu}$ with $0<\mu<2$. In this case,
the position of the walker after $m$ steps, grows as $m^{1/\mu}$ for large 
$m$~\cite{CM2005}. Hence the natural scaling limit is $x\to \infty$, $m\to 
\infty$ with the ratio $x/m^{1/\mu}$ fixed.
For $p_m(x)$, we expect a scaling form $p_m(x)\approx m^{-1/2-1/\mu} 
g_2(x/m^{1/\mu})$. The power of $m$ outside the scaling function is chosen
to ensure that 
$\int_0^{\infty} p_m(x)dx \sim m^{-1/2}$. This is needed since we know
from Eq. (\ref{q0}) and the Sparre Andersen theorem in Eq. (\ref{sa1})
that 
$\int_0^{\infty} p_m(x)dx=q_m(0)\sim 1/\sqrt{\pi m}$ for large $m$.
Similarly, for $q_m(x)$, we expect a scaling form
$q_m(x)\approx h_2(x/m^{1/\mu})$ in the scaling limit.  

To extract the leading scaling functions $g_2(z)$ and $h_2(z)$  
respectively from Eqs. (\ref{pm.2}) and (\ref{qm.2}), we need to
investigate the function $\phi(s,\lambda)$
in Eq. (\ref{ivanov.2})
in the corresponding scaling limit $\lambda\to 0$, $s\to 0$ but keeping
the ratio $\lambda/(1-s)^{1/\mu}$ fixed.
Fortunately, this was already done in Ref. 
\cite{CM2005} in a different context. Setting $s=1-p$ with $p\to 0$, the leading 
behavior
of $\phi(s,\lambda)$ in the scaling limit is given by (see Eqs. (43)-(47)
of Ref. \cite{CM2005})
\begin{equation}
\phi(s,\lambda)\approx \frac{1}{\sqrt{p}}\, 
\exp\left[-\frac{1}{\pi}\,\int_0^{\infty} \frac{du}{1+u^2}\, 
\ln\left[1+\frac{1}{p}\,(a\,\lambda\,u)^{\mu}\right]\right].
\label{phisl3}
\end{equation}

Let us first consider the function $p_m(x)$ in Eq. (\ref{pm.2}). We 
substitute the anticipated scaling form $p_m(x)= m^{-1/2-1/\mu}g_2(x/m^{1/mu})$
on the lhs of Eq. (\ref{pm.2}). As before, setting $p=1-s$, we can replace, in 
the scaling limit, the sum over $m$ by a continuous integral over $m$
\begin{equation}
\sum_{m=0}^{\infty}s^m\, \int_0^{\infty} p_m(x)\, e^{-\lambda x}\, dx
\approx \int_0^\infty \int_0^\infty dx\, dm e^{-\lambda x-p\, m} m^{-1/2-1/\mu}\, g_2(x m^{-1/\mu}).
\label{lhs.1}
\end{equation}
We then make a change of variable $x\,m^{-1/\mu}=z$ and $p\,m=y$ to get
\begin{equation}
\sum_{m=0}^{\infty}s^m\, \int_0^{\infty} p_m(x)\, e^{-\lambda x}\, dx
\approx \frac{1}{\sqrt{p}}\,\int_0^\infty \int_0^\infty dz\,dy\, g_2(z)\, y^{-1/2}\, e^{-(\lambda\, 
p^{-1/\mu})\, 
y^{1/\mu}\,z- y} 
\label{lhs.2}
\end{equation}
We next substitute Eq. (\ref{lhs.2}) on the lhs of Eq. (\ref{pm.2})
and Eq. (\ref{phisl3}) on the rhs of Eq. (\ref{pm.2}).
Writing the scaled variable as $\lambda\, p^{-1/\mu}=w$ and comparing 
lhs with the rhs, we see that the $1/\sqrt{p}$ cancels 
from both sides leaving us with 
\begin{equation}
\int_0^{\infty} dz\, g_2(z)\, \int_0^{\infty}dy\, y^{-1/2}\, e^{-y}\, e^{-w 
\,y^{1/\mu}\,z} = \exp\left[-\frac{1}{\pi}\int_0^{\infty} \frac{du}{1+u^2}\ln 
\left[1+ a^\mu w^{\mu} u^{\mu}\right]\right]\equiv J_\mu(w) \;.
\label{g2s.1}
\end{equation}
Similarly, by substituting the anticipated scaling form $q_m(x)= 
h_2(x/m^{1/\mu})$ on the lhs of Eq. (\ref{qm.2}) and doing exactly the same
series of manipulations, we get
\begin{equation}
\int_0^{\infty} dz\, h_2(z)\, \int_0^{\infty}dy\, y^{1/\mu}\, e^{-y}\, e^{-w
\,y^{1/\mu}\,z} = \frac{1}{w}\,J_\mu(w)
\label{h2s.1}
\end{equation}
where $J_\mu(w)$ is defined in Eq. (\ref{g2s.1}). 

For later purposes, it is further convenient to define a pair of Laplace 
transforms
\begin{eqnarray}
{\tilde g_2}(\rho) &=& \int_0^{\infty} g_2(z)\, e^{-\rho\, z}\, dz \label{g2_lap} \\
{\tilde h_2}(\rho) & =& \int_0^{\infty} h_2(z)\, e^{-\rho\, z}\, dz 
\label{h2_lap}
\end{eqnarray}
in terms of which Eqs. (\ref{g2s.1}) and (\ref{h2s.1}) read
\begin{eqnarray}
\int_0^{\infty} dy\, y^{-1/2}\, e^{-y}\, {\tilde g_2}(w\,y^{1/\mu}) &= & J_\mu(w) 
\label{g21.lap}\\
\int_0^{\infty} dy\, y^{1/\mu}\, e^{-y}\, {\tilde h_2}(w\,y^{1/\mu}) &=& 
\frac{1}{w}\,J_\mu(w)
\label{h21.lap}
\end{eqnarray}

The equations (\ref{g2s.1}) and (\ref{h2s.1}) determine, in principle,
the two scalings functions $g_2(z)$ and $h_2(z)$ for all $z$. In practice, 
it is 
hard to
invert these two equations to obtain $g_2(z)$ and $h_2(z)$ for all $z$.
However, it is possible to extract the large $z$ asymptotics of these two
functions by analyzing the leading singular behavior of $J_\mu(w)$ 
in Eq. (\ref{g2s.1}) as $w\to 0$. 
Clearly, it follows from the definition of $J_\mu(w)$ in Eq. (\ref{g2s.1})
that $J_\mu(0)=1$. We are, however, interested in the leading singular correction 
term
in $J_\mu(w)$ as $w\to 0$ which, as it turns out, depends 
on whether $0<\mu<1$, $1<\mu<2$ or $\mu=1$. Below, we consider the three 
cases separately.

\subsection{The case $0<\mu<1$}

We consider $J_\mu(w)$ in Eq. (\ref{g2s.1})
and compute the derivative $J_\mu'(w)$ as $w\to 0$. Simple computation shows that
\begin{equation}
J_\mu'(w)\xrightarrow{w\to 0} - \mu\, b_\mu w^{\mu-1}; \quad {\rm where}\quad 
b_\mu= \frac{a^\mu}{\pi} \int_0^{\infty} \frac{u^\mu\, du}{1+u^2}.
\label{bmudef.1}
\end{equation}
Note that the integral defining $b_\mu$ is convergent as $u\to \infty$ for
$0<\mu<1$.
Integrating over $w$ and using $J_\mu(0)=1$ we get the leading correction term
as $w\to 0$
\begin{equation}
J_\mu(w) \approx 1 - b_\mu w^{\mu}+\ldots  
\label{jmu01}
\end{equation}
where $b_\mu$ is given in Eq. (\ref{bmudef.1}).

Substituting this small $w$ behavior of $J_\mu(w)$ on the rhs of Eq. 
(\ref{g21.lap}), it follows that to match the powers of $w$ on both sides,
the Laplace transform ${\tilde g_2}(\rho)$ must have the following
small $\rho$ behavior
\begin{equation}
{\tilde g_2}(\rho)\underset{\rho\to 0}{\sim} \frac{1}{\sqrt{\pi}}- 
\frac{2}{\sqrt{\pi}}\,b_\mu\, \rho^\mu \,.
\label{g22.lap}
\end{equation}
Using the classical Tauberian theorem (for a simple derivation see the appendix
A.2 of Ref. \cite{MEZ2006}), one immediately gets the following large $z$ 
behavior of $g_2(z)$
\begin{equation}
g_2(z)\underset{z\to \infty}{\sim} \frac{A_\mu}{z^{1+\mu}}
\label{g22.largez}
\end{equation}
with the amplitude 
\begin{equation}
A_\mu= \frac{2\mu}{\sqrt{\pi}}\,\frac{b_\mu}{\Gamma(1-\mu)}= 
\frac{2\mu}{\sqrt{\pi}}\,\beta_\mu
\quad {\rm where} \quad \beta_\mu= \frac{b_\mu}{\Gamma(1-\mu)}= 
\frac{a^{\mu}}{\pi
\Gamma(1-\mu)}\,\int_0^{\infty}\frac{u^{\mu}}{1+u^2}\, du\,. 
\label{amudef}
\end{equation}

Similarly, substituting the small $w$ behavior of $J_\mu(w)$ on the rhs of
Eq. (\ref{h21.lap}) and matching powers of $w$ on both sides, we get 
\begin{equation}
{\tilde h_2}(\rho)\underset{\rho\to 0}{\sim}\frac{1}{\rho}- b_{\mu}\,\rho^{\mu-1}\, .
\label{h22.lap}
\end{equation}
Once again, using the Tauberian theorem of inversion, we get
\begin{equation}
h_2(z)\underset{z\to \infty}{\sim} 1- \frac{B_\mu}{z^\mu}
\label{h22.largez}
\end{equation}
with the amplitude
\begin{equation}
B_\mu = \frac{b_\mu}{\Gamma(1-\mu)}= \beta_\mu \;,
\label{bmudef}
\end{equation}
where $\beta_\mu$ is given in Eq. (\ref{amudef}).

Finally, note that the ratio
\begin{equation}
\frac{A_\mu}{\mu\,B_\mu}= \frac{2}{\sqrt \pi}
\label{ratio.01}
\end{equation}
is universal in the sense that it is independent of $\mu \in (0,1)$ as well as 
on the scale factor $a$.

\subsection{The case $1< \mu<2$}

Unlike in the previous case, one finds that
the first derivative of $J_\mu(w)$ at $w=0$ is finite when $\mu\in [0,2]$
and is given by
\begin{equation}
\alpha_\mu=J_\mu'(0)= -\frac{a\,\mu }{\pi}\, \int_0^{\infty} \frac{z^{\mu-2}\, 
dz}{1+z^\mu}.
\label{jmuder.1}
\end{equation}
Note that for $1<\mu<2$, the integral in Eq. (\ref{jmuder.1}) is 
convergent
as $z\to 0$.
Thus, as $w\to 0$, $J_\mu(w)\to 1- \alpha_\mu\, w$. To obtain the leading
non-analytic singular term, we need to compute the next term. By taking
two derivatives with respect to $w$ near $w=0$ and then re-integrating back,
we find the following leading singular behavior of $J_\mu(w)$ near $w=0$
\begin{equation}
J_\mu(w)\approx 1- \alpha_\mu\, w + c_\mu \, w^{\mu}+\ldots \quad {\rm 
where}\quad 
c_\mu= \frac{2a^\mu}{\pi(\mu-1)}\,\int_0^{\infty} \frac{u^\mu\, du}{(1+u^2)^2}.
\label{jmu12}
\end{equation}
   
Substituting this small $w$ behavior of $J_\mu(w)$ on the rhs of Eq. 
(\ref{g21.lap}) and matching powers of $w$ on both sides we get
\begin{equation}
{\tilde g_2}(\rho)\underset{\rho\to 0}{\sim} \frac{1}{\sqrt{\pi}}- 
\frac{\alpha_\mu}{\Gamma(1/2+1/\mu)}\, \rho + \frac{2}{\sqrt{\pi}}\, c_\mu\, 
\rho^{\mu}
\label{g22.lap12}
\end{equation}
where $\alpha_\mu$ and $c_\mu$ are given respectively in Eqs. (\ref{jmuder.1})
and (\ref{jmu12}).
Again, inverting via the Tauberian theorem (see Ref. \cite{MEZ2006}), we 
get
\begin{equation}
g_2(z)\underset{z\to \infty}{\sim} \frac{A_\mu}{z^{1+\mu}} \;,
\label{g22.largez12}
\end{equation}
with the amplitude
\begin{equation}
A_\mu=\frac{2}{\sqrt{\pi}}\, 
\frac{\mu(\mu-1)c_\mu}{\Gamma(2-\mu)}=\frac{2\mu}{\sqrt{\pi}}\,\beta_\mu
\quad {\rm where} \quad \beta_\mu=\frac{2a^\mu}{\pi 
\Gamma(2-\mu)}\,\int_0^{\infty}
\frac{u^{\mu}}{(1+u^2)^2}\, du \,.
\label{amudef12}
\end{equation}

Exactly in a similar way, we substitute the small $w$ behavior of $J_\mu(w)$
on the rhs of Eq. (\ref{h21.lap}), match powers of $w$ on both sides  
and find that
\begin{equation}
{\tilde h_2}(\rho)\underset{\rho\to 
0}{\sim}\frac{1}{\rho}-\frac{\alpha_\mu}{\Gamma(1+2/\mu)} + c_\mu\, \rho^{\mu-1}
\label{h22.lap12}
\end{equation}
where $\alpha_\mu$ and $c_\mu$ are defined in Eqs. (\ref{jmuder.1})
and (\ref{jmu12}). Inverting via the Tauberian theorem gives
the desired result
\begin{equation}
h_2(z)\underset{z\to \infty}{\sim} 1- \frac{B_\mu}{z^\mu}
\label{h22.largez12}
\end{equation}
with the amplitude
\begin{equation}
B_\mu = \frac{(\mu-1)c_\mu}{\Gamma(2-\mu)}= \beta_\mu
\label{bmudef12}
\end{equation}
where $\beta_\mu$ is given in Eq. (\ref{amudef12}).

In this case, also we note that the ratio
\begin{equation}
\frac{A_\mu}{\mu\,B_\mu}= \frac{2}{\sqrt \pi}
\label{ratio.12}
\end{equation}
is universal and does not depend explicitly on $1<\mu<2$ and $a$.

\subsection{The case $\mu=1$}

In this case, from Eq. (\ref{g2s.1})
\begin{equation}
J_1(w)= \exp[-I_1(w)] \quad {\rm where}\quad I_1(w)= \frac{1}{\pi}\int_0^{\infty} 
\frac{du}{1+u^2}\, \ln (1+a\,w\,u) \;.
\label{j1w.1}
\end{equation}
Let us first derive the leading singular behavior of $I_1(w)$ as $w\to 0$. Making 
a change of variable $x=a\,w\,u$ in the integral we get
\begin{equation}
I_1(w)= \frac{aw}{\pi}\int_0^{\infty} \frac{dx}{x^2+a^2w^2}\, \ln(1+x) \;.
\label{i1w.1}
\end{equation}
Next, we divide the integration range $[0,\infty)$
into two parts $[0,1]$ and $[1,\infty)$
and write $I_1(w)= Z_1(w)+ Z_2(w)$.
The second part $Z_2(w)$, i.e., the integral over 
$[1,\infty]$
is a completely analytic function as $w\to 0$. Thus the leading singular
behavior of $I_1(w)$ as $w\to 0$ is contained only in the first part
\begin{equation}
Z_1(w) = \frac{aw}{\pi}\int_0^1 \frac{dx}{x^2+a^2w^2}\, \ln(1+x) \;.
\label{z1w.1}
\end{equation}
In this integral, we can now safely expand $\ln(1+x)= x-x^2/+x^3/3+\ldots$
and perform the integral term by term. The leading singularity comes
from the first term of this expansion
\begin{equation}
Z_1(w)\approx \frac{aw}{\pi}\int_0^1 \frac{x}{x^2+a^2w^2}\, dx= 
\frac{aw}{\pi}\ln\left[\frac{\sqrt{1+a^2w^2}}{aw}\right]\underset{w\to 0}{\sim}- 
\frac{a}{\pi}\,w\, \ln w
\label{z1w.2}
\end{equation}
which indicates, from Eq. (\ref{j1w.1}), that 
\begin{equation}
J_1(w)\underset{w\to 0}{\sim} 1 + \frac{a}{\pi}\, w\, \ln w \;.
\label{j1w.2}
\end{equation}

Substituting this small $w$ behavior of $J_1(w)$ on the rhs of Eq.
(\ref{g21.lap}) and matching the leading behavior of $w$ on both sides
indicates that
\begin{equation}
{\tilde g_2}(\rho)\underset{\rho\to 0}{\sim} \frac{1}{\sqrt{\pi}} + 
\frac{2}{\sqrt{\pi}}\, \frac{a}{\pi}\, \rho\, \ln \rho \, .
\label{g22.lap1}
\end{equation}
This indicates, using Tauberian inversion theorem (see Ref. \cite{MEZ2006}),
\begin{equation}
g_2(z)\underset{z\to \infty}{\sim} \frac{A_1}{z^2}\quad {\rm where}\quad 
A_1=\frac{2}{\sqrt{\pi}}\, \frac{a}{\pi}.
\label{g22.largez1}
\end{equation}

Similarly, substituting the small $w$ behavior of $J_\mu(w)$ on the rhs of
Eq. (\ref{h21.lap}) and matching leading behavior of $w$ on both sides
we get
\begin{equation}   
{\tilde h_2}(\rho)\underset{\rho\to 0}{\sim}\frac{1}{\rho}+ \frac{a}{\pi}\, \ln \rho
\label{h22.lap1}
\end{equation}
which, when inverted, provides the following large $z$ behavior 
\begin{equation}
h_2(z)\underset{z\to \infty}{\sim} 1- \frac{B_1}{z} \quad {\rm where} \quad 
B_1=\frac{a}{\pi} \;.
\label{h22.largez1}
\end{equation}
Finally, we notice that even for this marginal $\mu=1$ case, the ratio
$A_1/B_1=2/\sqrt{\pi}$ has the same value as in the other two cases, namely
for $0<\mu<1$ and $1<\mu<2$.

Let us remark that if one puts $\mu=1$ in the expression of $\beta_\mu$ in Eq. 
(\ref{betamu2}), we get $\beta_1=a/\pi$. Correspondingly $A_1=2a/{\pi}^{3/2}$
from Eq. (\ref{amu1}) and $B_1=a/\pi$ from Eq. (\ref{bmu1}), we find
that they are consistent respectively with $A_1$ in (\ref{g22.largez1})
and and $B_1$ in (\ref{h22.largez1}). In other words, the final
asymptotic results for $g_2(z)$ and $h_2(z)$ in
the marginal case $\mu=1$ are included in the range $\mu\in [1,2]$, even though
the details for $\mu=1$ are quite different, as it has logarithmic singularities.

\section{Distribution of the maximum of a lattice random walk}\label{app_lattice}

In this appendix we consider $N$ lattice random walks (RW) starting at $x_i(0) = 0$, for 
$i=1,2, \cdots, N$ and evolving as
\begin{eqnarray}
x_i(m) = x_i(m-1) + \eta_i(m) \;,
\end{eqnarray}
where the noise $\eta_i(m)$'s are i.i.d. random variables with a distribution 
$f(\eta) = \frac{1}{2} \delta(\eta-1) + \frac{1}{2} \delta(\eta+1)$. 
The aim is to show the result in Eq. (\ref{exact_Bernoulli}), 
taking advantage of the relation (\ref{identity}). 

We first consider a single random walk, $N=1$, and denote by $W(j,n)$ the number of lattice RW 
starting at $x_1(0)=0$ and ending in $j$ after $n$ steps. One has
\begin{eqnarray}\label{free_walk}
W(j,n) = 
\begin{cases}
&{n \choose k} \:, \; 2k = n + j  \;, \; n+j \; {\rm even} \;, \\
&0 \;, \; n+j \; {\rm odd} \;. 
\end{cases}
\end{eqnarray}
To compute the cumulative distribution function (cdf) of the maximal displacement of $N$ 
walkers we need to compute the number of walks, for a single walker $N=1$, which stay strictly below a given value $M$. We thus denote, for $N=1$, by $W_M(j,n)$ the number of walks which stays strictly below an integer $M$ and end up in $j$ after $n$ steps. To do this we use the reflection principle, e. g. the method of images: $W_M(j,n)$ can be obtained by subtracting to $W(j,n)$ the number of free walks which start in $x(0) = 2M$ and end in $j$ after $n$ steps. This yields: 
\begin{eqnarray}\label{reflection_1}
W_M(j,n) = \begin{cases}
&{n \choose k} -  {n \choose k-M}\:, \; 2k = n + j  \;, \; n+j \; {\rm even} \;, \\
&0 \;, \; n+j \; {\rm odd} \;. 
\end{cases}
\end{eqnarray}
The total number of walks $W_M(n)$ which start at $x_1(0)=0$ and stay strictly below $M$ 
after $n$ steps are obtained by summing $W(j,n)$ in Eq. (\ref{reflection_1}) 
over the endpoint $j < M$. This yields
\begin{eqnarray}\label{walks_box}
W_M(n) = \sum_{k=0}^{\lfloor \frac{n+M}{2}\rfloor} \left[{n \choose k} - {n \choose 
k-M}\right]  \;,
\end{eqnarray}
where $\lfloor x \rfloor$ is the largest integer not greater than $x$. Therefore one has
\begin{eqnarray}
{\rm Proba.}\, \left[\max_{0\leq m \leq n} x_1(m) < M \right] =
\frac{W_M(n)}{2^n} = \frac{1}{2^n} \sum_{k=0}^{\lfloor \frac{n+M}{2}\rfloor} \left[{n 
\choose k} - {n \choose k-M}\right]  \;.
\end{eqnarray}

We can now write the cdf of the maximal 
displacement of $N$ independent walkers as
\begin{eqnarray}
{\rm Proba.}\, \left[\max_{0\leq m \leq n} x_{\rm max}(m) < M \right] = 
\left(\frac{W_M(n)}{2^n}\right)^N = 
\left(\frac{1}{2^n} \sum_{k=0}^{\lfloor \frac{n+M}{2}\rfloor} \left[ {n \choose k} - {n 
\choose k-M}\right]\right)^N  \;,
\end{eqnarray}
where $x_{\rm max}(m) = \max_{1 \leq i \leq N} x_i(m)$, from which one gets
\begin{eqnarray}
{\rm Proba.} \left[\max_{0\leq m \leq n} x_{\max}(m)  = M \right] = \frac{1}{2^{nN}} \left([W_{M+1}(n)]^N - [W_M(n)]^N \right) \;.
\end{eqnarray}
Finally, using the identity (\ref{identity}), 
one obtains the result given in the text in Eq. (\ref{exact_Bernoulli}).

\end{document}